\batchmode
\makeatletter
\def\input@path{{\string"C:/Users/marfon/OneDrive - The University of Liverpool/Texts/ClusteredPMBM/\string"}}
\makeatother
\documentclass[british]{IEEEtran}
\usepackage[T1]{fontenc}
\usepackage[latin9]{inputenc}
\usepackage{float}
\usepackage{amsmath}
\usepackage{amsthm}
\usepackage{amssymb}
\usepackage{graphicx}

\makeatletter

\providecommand{\tabularnewline}{\\}
\floatstyle{ruled}
\newfloat{algorithm}{tbp}{loa}
\providecommand{\algorithmname}{Algorithm}
\floatname{algorithm}{\protect\algorithmname}
\usepackage{tikz}
\usetikzlibrary{calc}

\theoremstyle{plain}
\newtheorem{thm}{\protect\theoremname}
\theoremstyle{definition}
\newtheorem{defn}[thm]{\protect\definitionname}
\theoremstyle{plain}
\newtheorem{lem}[thm]{\protect\lemmaname}

\usepackage{cite} 
\usepackage[margin=8pt,font=footnotesize]{caption}
\usepackage{algorithm}
\newcommand{\StatexIndent}[1][3]{%
  \setlength\@tempdima{\algorithmicindent}%
  \Statex\hskip\dimexpr#1\@tempdima\relax}
\usepackage{algpseudocode}
\algdef{S}[FORALL]{ForAllNoDo}[1]{\algorithmicforall #1}%
\algdef{S}[FOR]{ForNoDo}[1]{\algorithmicfor\ #1}%

\usepackage{amsmath}  
\allowdisplaybreaks

\makeatother

\usepackage{babel}
\providecommand{\definitionname}{Definition}
\providecommand{\lemmaname}{Lemma}
\providecommand{\theoremname}{Theorem}

\begin{document}

\title{Data-driven clustering and Bernoulli merging for the Poisson multi-Bernoulli
mixture filter}

\author{Marco Fontana, Ángel F. García-Fernández, Simon Maskell\thanks{M. Fontana, A. F. García-Fernández and S. Maskell are with the Department of Electrical Engineering and Electronics, University of Liverpool, Liverpool L69 3GJ, United Kingdom (emails: \{marco.fontana, angel.garcia-fernandez, s.maskell\}@liverpool.ac.uk). A. F. García-Fernández is also with the ARIES Research Centre, Universidad Antonio de Nebrija,  Madrid, Spain.}}
\maketitle
\begin{abstract}
This paper proposes a clustering and merging approach for the Poisson
multi-Bernoulli mixture (PMBM) filter to lower its computational complexity
and make it suitable for multiple target tracking with a high number
of targets. We define a measurement-driven clustering algorithm to
reduce the data association problem into several subproblems, and
we provide the derivation of the resulting clustered PMBM posterior
density via Kullback-Leibler divergence minimisation. Furthermore,
we investigate different strategies to reduce the number of single
target hypotheses by approximating the posterior via merging and inter-track
swapping of Bernoulli components. We evaluate the performance of the
proposed algorithm on simulated tracking scenarios with more than
one thousand targets.
\end{abstract}


\section{Introduction\label{sec:Introduction}}

Multi-target tracking (MTT) is a well-known problem of interest in
many application fields, including surveillance, traffic control and
autonomous driving \cite{Vo15,Chang2010,Meyer2018}. The main goal
is the estimation of the number of targets and their states based
on the noisy measurements recorded by a sensor, which includes false
alarms and missed detections. The targets move in a dynamic scenario,
appearing and disappearing from the field of view of the sensor.

MTT has been studied for decades, and several solutions have been
proposed to improve the trade-off between performance and computational
efficiency. Among the most widely-used approaches, we mention multiple
hypothesis tracking (MHT) \cite{Reid79,Kurien90,Blackman91,Brekke18,Coraluppi2014},
joint probabilistic data association (JPDA) \cite{Fortmann83} and
random finite sets (RFS) \cite{Mahler14}. 

In the last decade, several solutions to the MTT problem have been
based on different birth models. With Poisson point process (PPP)
birth model and the standard measurement/dynamic models, the posterior
is a Poisson multi-Bernoulli mixture (PMBM) \cite{Williams15a,Garcia-Fernandez18}.
With multi-Bernoulli birth, the conjugate prior is a multi-Bernoulli
mixture (MBM), which can be labelled and written in $\delta$-generalised
labelled multi-Bernoulli ($\delta$-GLMB) form \cite[Sec. IV]{Garcia-Fernandez18}\cite{Vo2014}.
Approximate filters based on the PMBM and $\delta$-GLMB filters are
the Poisson multi-Bernoulli (PMB) filters \cite{Williams12,Williams15}
and the labelled multi-Bernoulli (LMB) filter \cite{Reuter2014}.

The PMBM filter can be considered a state-of-the-art fully Bayesian
MHT filter, with an efficient representation of global hypotheses
with probabilistic target existence (Bernoulli components) and information
on undetected targets. The (track-oriented) PMB filter, which only
propagates one component of the PMBM filter, can be seen as a fully
Bayesian version of the integrated JPDA filter, with differences explained
in \cite[Section IV. A]{Williams15a}.

Of particular importance in real-world scenarios is to be able to
track a large number of targets.  The main challenge in large-scale
MTT problems is the evaluation of all the possible measurement-target
associations, which is known as data association problem. Several
approaches were developed to efficiently manage the large number of
hypotheses resulting from this combinatorial problem \cite{Maskell04,Collins1992,Musicki2008,Wang1999,Campbell2021}.
Among several methods, clustering and hypothesis merging are some
of the most popular and most effective solutions. We briefly review
the literature on these topics.

Clustering is considered one of the most effective strategies to increase
the scalability of tracking algorithms. In a context of sufficiently-sparse
targets, clustering addresses large data association problems defining
several independent subproblems. In the original MHT paper \cite{Reid79},
the author describes a procedure to associate measurements with clusters
of independent potential targets, merging the clusters associated
to common measurements, and creating new individual clusters for the
measurements not associated with any potential target in the prior.
This method, as the following ones derived from \cite{Reid79}, can
be referred to hypothesis clustering \cite{Drummond1993}. Similar
procedures have been used in \cite{Werthmann1992} to cluster targets
based on the data association in multiple scans. In \cite{Uhlmann90}
the authors propose a spatial clustering of the tracks based on a
minimum separation distance, addressing the assignment of new measurements
to the most appropriate cluster using the gating procedure. In \cite{Waard2001}
a review of the split method presented in \cite{Kurien90} is used
to initialise new clusters based on the independent components in
each cluster, detected by using rectangular areas in the measurement
space. In \cite{Roy97} the authors describe an efficient cluster
management approach based on a dynamic data structure, which implements
the hypothesis tree. These methods do not have a straightforward application
on multi-Bernoulli filters, as they are based on the concept of confirmed
target, which is not defined for targets with probabilistic target
existence.

For the $\delta$-GLMB filter, a clustering algorithm for large-scale
tracking based on predicted measurements gating regions is proposed
in \cite{Beard20}. A drawback of this approach is that the $\delta$-GLMB
representation of a labelled MBM involves an exponential increase
in the number of global hypothesis, which requires extra computational
time \cite{Garcia-Fernandez18}. In addition, this implementation
neglects information on undetected targets, which is required in fully
Bayesian MTT \cite{GarciaFernandez2022a} and important in many applications,
for example, autonomous vehicles and search-and-track \cite{BostroemRost2021}.

Merging is another popular approach used to decrease the number of
hypotheses in the filters and therefore computational time. One of
the first contributions of this kind can be found in \cite{Singer1974},
where the authors\textit{ }suggest merging all tracks which share
measurements for the past $N$ times.

The first approaches to improve the computational efficiency of the
PMBM filter were proposed in \cite{Garcia-Fernandez18}, where the
authors suggested to cap the number of global hypotheses in the filter
and pruning the Poisson and Bernoulli components whose weight is below
a threshold. An alternative approach is to perform track-oriented
N-scan pruning \cite{Xia2020}. Two efficient PMB approximations of
the PMBM posterior density were proposed in \cite{Xia2022} for multiple
extended object filtering, based on the original paper \cite{Williams15a}
and the variational approximation via Kullback--Leibler divergence
minimisation.

\begin{figure}
\centering{}\includegraphics[scale=0.4]{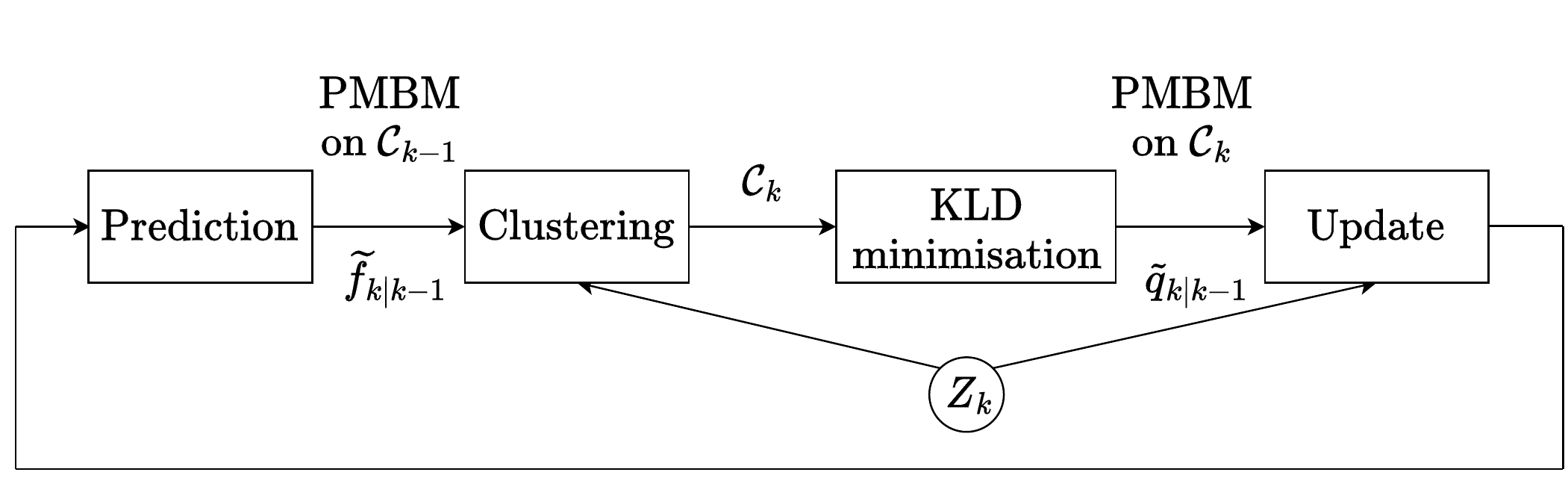}\caption{Schematic of the clustered PMBM filter. The prediction step propagates
a clustered PMBM density on the set of clusters $\mathcal{C}_{k-1}$
computed at the previous time instant. Once the filter receives the
set of measurements $Z_{k}$ for the current time instant $k$, we
obtain the new set of clusters $\mathcal{C}_{k}$ and perform a KLD
minimisation (with auxiliary variables) to obtain the current clustered
PMBM. The clustering algorithm can be defined by the user. \label{fig:filter-schematic}}
\end{figure}

In this work, we focus on developing novel clustering and merging
algorithms to provide an efficient implementation of the PMBM filter.
We proceed to explain the contributions. Our first contribution is
a new data-driven clustering technique with low computational burden
for the PMBM filter. To the best of our knowledge, this is the first
attempt to cluster Bernoulli components in a PMBM filter implementation.
We first define the concept of clustered PMBM density that is of general
validity for any clustering algorithm. A clustered PMBM density is
the union of independent a PPP and a number of independent MBMs, one
for each cluster. We obtain the best fitting clustered PMBM by minimising
the Kullback-Leibler divergence (KLD) after introducing auxiliary
variables over the track indices in the target space \cite{GarciaFernandez2020a},
see diagram in Fig. \ref{fig:filter-schematic}. Then, the proposed
clustering algorithm takes into account the predicted density and
the received measurements to group potential targets that may have
given rise to a common measurement, which is computationally advantageous
compared to spatial clustering techniques.

Our second contribution is a Bernoulli merging approach for local
hypotheses corresponding to the same potential target. We compute
the similarity between Bernoulli components via the KLD and merge
the most similar ones, as presented in \cite{Fontana2020}. The proposed
method allow us to obtain an accurate representation of the filter
posterior, merging similar local hypotheses with different data association
history.

Our third contribution aims to improve the efficiency of the clustering
algorithm for situations in which targets move in close proximity
and then separate. In this setting, some Bernoulli components of different
potential targets may overlap even though the actual locations of
the potential targets are already well separated, which hinders clustering.
To address this, we propose to swap certain Bernoulli components of
different potential targets, keeping the PMBM representation unaltered,
so that all the Bernoulli components of the same potential target
are found in the same region, and clustering can be done efficiently.
This approach bears resemblance to the particle swapping approach
used in the particle filter for track-before-detect in \cite{Kreucher05}
and also to the set JPDA algorithm \cite{Svensson2011} and variational
PMB filters \cite{Williams15}.

The paper is organised as follows. Background on the PMBM filter is
provided in Section \ref{sec:Background}. In Section \ref{sec:Clustering_KLD},
given the clusters of potential targets, we provide the clustered
posterior density of the filter through KLD minimisation. The measurement-driven
clustering algorithm is presented in Section \ref{sec:Clustering_Measurements}.
Section \ref{sec:Merging} introduces two strategies to decrease the
number of Bernoulli components via intra-track and inter-track Bernoulli
merging. Finally, we evaluate filter performance via simulations in
Section \ref{sec:Simulations}, and we draw the conclusions in Section
\ref{sec:Conclusions}.

\section{Background: PMBM filtering\label{sec:Background}}

In this section, we briefly review the standard dynamic model and
the standard point target measurement model in Section \ref{subsec:Bayesian-Filtering-Recursion}.
Section \ref{subsec:PMBM_Intro} provides an overview of the PMBM
filter; for a more extensive description we refer the reader to \cite{Williams15a,Garcia-Fernandez18}.
Finally, we introduce auxiliary variables in the PMBM in Section \ref{subsec:auxiliary variables}.

\subsection{Multi-target system modelling\label{subsec:Bayesian-Filtering-Recursion}}

In the context of multi-target systems, we regard filtering as the
estimation of the states of a time-varying number of targets at the
current time step $k$. We denote a single target state as $x_{k}\in\mathbb{R}^{n_{x}}$,
and the set of target states at time $k$ as $X_{k}\in\mathcal{F}(\mathbb{R}^{n_{x}})$,
where $\mathcal{F}(\mathbb{R}^{n_{x}})$ is the set of all finite
subsets of $\mathbb{R}^{n_{x}}$. The set $X_{k}$ is modelled as
a RFS, meaning that both the cardinality $|X_{k}|$ and the elements
of the set (i.e., the target states) are random variables \cite{Mahler14}.

At each time step, a target with state $x$ survives with probability
$p_{S}(x)$, or departs with probability $1-p_{S}(x)$ independently
of the rest of the targets. The evolution of a surviving target can
be defined by a Markov transition density $g(\cdot|x)$. At time step
$k+1$, the multi-target state $X_{k+1}$ is the union of the surviving
targets and the independent new targets, which are modelled by a PPP
with intensity $\lambda(\cdot)$.

The set of measurements at time $k$ is denoted by $Z_{k}\in\mathcal{F}(\mathbb{R}^{n_{z}})$
and is the union of target-generated measurements and independent
PPP clutter with intensity $\lambda_{C}(\cdot)$. At each time step,
an existing target $x_{k}$ is detected with probability $p_{D}(x_{k})$,
or misdetected with probability $1-p_{D}(x_{k})$. Each detected target
$x_{k}\in X_{k}$ generates a measurement $z_{k}$ with density $l(z_{k}|x_{k})$.

\subsection{PMBM density\label{subsec:PMBM_Intro}}

For the models described in Section \ref{subsec:Bayesian-Filtering-Recursion},
the density $f_{k'|k}(\cdot)$ of the set of targets at time step
$k'\in\left\{ k,k+1\right\} $ given measurements up to time step
$k$ is a PMBM \cite{Williams15a}. That is, it results from the union
of two independent RFSs: a PPP with density $f_{k'|k}^{p}(\cdot)$,
and a MBM RFS with density $f_{k'|k}^{mbm}(\cdot)$. The PMBM density
is expressed as
\begin{equation}
f_{k'|k}(X_{k'|k})=\sum_{Y\uplus W=X_{k'|k}}f_{k'|k}^{p}(Y)f_{k'|k}^{mbm}(W)\label{eq:PMBM_density}
\end{equation}
where the sum goes over all mutually disjoint sets $Y$ and $W$,
such that their union is $X_{k'|k}$.

The PPP density represents the targets that exist at the current time,
but have not yet been detected. Its density is
\begin{equation}
f_{k'|k}^{p}(X)=e^{-\int\lambda_{k'|k}(x)dx}\prod_{x\in X}\lambda_{k'|k}(x)\label{eq:PPP}
\end{equation}
where $\lambda_{k'|k}(\cdot)$ is the intensity. In the PPP, the cardinality
is Poisson distributed and targets are independent, and identically
distributed. The MBM part represents the potentially detected targets,
and it can be described as \cite{Williams15a}
\begin{equation}
f_{k'|k}^{mbm}(X)=\sum_{a\in\mathcal{A}_{k'|k}}w_{_{k'|k}}^{a}\sum_{\uplus_{j=1}^{n_{k'|k}}X^{j}=X}\thinspace\prod_{i=1}^{n_{k'|k}}f_{k'|k}^{i,a^{i}}(X^{i})\label{eq:MBM}
\end{equation}
where $i$ is the index over the Bernoulli components, $a=(a^{1},\dots,a^{n_{k'|k}})\in\mathcal{A}_{k'|k}$
represents a specific data association hypothesis, $a^{i}\in\{1,\dots,h_{k'|k}^{i}\}$
is an index over the $h_{k'|k}^{i}$ single target hypotheses for
the $i$-th potential target, and $n_{k'|k}$ is the number of potentially
detected targets. Each set of single target hypothesis $a\in\mathcal{A}_{k'|k}$
is also called a global hypothesis (whose mathematical expression
is provided in \cite{Williams15a}), and it is associated to a weight
$w_{_{k'|k}}^{a}$ satisfying $\sum_{a\in\mathcal{A}_{k'|k}}w_{_{k'|k}}^{a}=1$.

The Bernoulli density corresponding to the $i$-th potential target,
$i\in\{1,\dots n_{k'|k}\},$ and the $a_{i}$ single target hypothesis
density $f_{k'|k}^{i,a^{i}}(X)$ can describe a newly detected target,
a previously detected target or clutter. It efficiently models both
the uncertainty regarding target existence and state. Mathematically,
it can be expressed as
\begin{align}
f_{k'|k}^{i,a_{i}}\left(X\right) & =\begin{cases}
1-r_{k'|k}^{i,a^{i}} & X=\emptyset\\
r_{k'|k}^{i,a^{i}}p_{k'|k}^{i,a^{i}}(x) & X=\left\{ x\right\} \\
0 & \mathrm{otherwise}
\end{cases}\label{eq:Bernoulli_density_filter}
\end{align}
where $r_{k'|k}^{i,a^{i}}\in[0,1]$ is the probability of existence
and $p_{k'|k}^{i,a^{i}}(\cdot)$ is the state density given that it
exists. We often refer to a potential target as a track, which is
defined as a collection of single target hypotheses corresponding
to the same potential target \cite{Williams15a}.

The prediction and update steps of the PMBM filter to obtain (\ref{eq:PMBM_density})
are given in \cite{Williams15a,Garcia-Fernandez18}.

\subsection{PMBM with auxiliary variables\label{subsec:auxiliary variables}}

In order to perform clustering on the PMBM density (\ref{eq:PMBM_density})
based on KLD minimisation, which will be done in Section \ref{sec:Clustering_KLD},
we require to introduce auxiliary variables in the density (\ref{eq:PMBM_density}),
as done in \cite{Williams15a,GarciaFernandez2020a} for PMB filters.
Auxiliary variables do not change the PMBM/MBM distribution, they
just make implicit information in the posterior explicit. Similar
approaches to introduce auxiliary/hidden variables in mixtures of
densities can be found in the particle filtering and expectation maximisation
literature \cite{Pitt99,Bishop_book06}.

Given (\ref{eq:PMBM_density}), the target state space is augmented
with an auxiliary variable $u\in\mathbb{\mathbb{U}}_{k'|k}=\left\{ 0,1,..,n_{k'|k}\right\} $,
such that $\left(u,x\right)\in\mathbb{\mathbb{U}}_{k'|k}\times\mathbb{R}^{n_{x}}$.
Variable $u=0$ implies that the target has not yet been detected,
so it corresponds to the PPP, and $u=i$ indicates that the target
corresponds to the $i$-th Bernoulli component. We denote a set of
target states with auxiliary variables as $\widetilde{X}_{k'}\in\mathcal{F}\left(\mathbb{\mathbb{U}}_{k'|k}\times\mathbb{R}^{n_{x}}\right)$.
\begin{defn}
\label{def:aux_density}Given $f_{k'|k}\left(\cdot\right)$ of the
form (\ref{eq:PMBM_density}), the density $\widetilde{f}_{k'|k}\left(\cdot\right)$
on the space $\mathcal{F}\left(\mathbb{\mathbb{U}}_{k'|k}\times\mathbb{R}^{n_{x}}\right)$
of sets of target states with auxiliary variable is \cite{GarciaFernandez2020a}
\begin{align}
 & \widetilde{f}_{k'|k}\left(\widetilde{X}_{k'}\right)\nonumber \\
 & =\sum_{\uplus_{l=1}^{n_{k'|k}}\widetilde{X}^{l}\uplus\widetilde{Y}=\widetilde{X}_{k'}}\widetilde{f}_{k'|k}^{p}\left(\widetilde{Y}\right)\sum_{a\in\mathcal{A}_{k'|k}}w_{k'|k}^{a}\prod_{i=1}^{n_{k'|k}}\left[\widetilde{f}_{k'|k}^{i,a^{i}}\left(\widetilde{X}^{i}\right)\right]\nonumber \\
 & =\widetilde{f}_{k'|k}^{p}\left(\widetilde{Y}_{k'}\right)\sum_{a\in\mathcal{A}_{k'|k}}w_{k'|k}^{a}\prod_{i=1}^{n_{k'|k}}\left[\widetilde{f}_{k'|k}^{i,a^{i}}\left(\widetilde{X}_{k'}^{i}\right)\right]\label{eq:PMBM_aux_var2}
\end{align}
where, for a given $\widetilde{X}_{k'}$, $\widetilde{Y}_{k'}=\left\{ \left(u,x\right)\in\widetilde{X}_{k'}|u=0\right\} $
and $\widetilde{X}_{k'}^{i}=\left\{ \left(u,x\right)\in\widetilde{X}_{k'}|u=i\right\} $,
and
\begin{align}
\widetilde{f}_{k'|k}^{p}\left(\widetilde{X}_{k'}\right) & =e^{-\int\lambda_{k'|k}\left(x\right)dx}\prod_{(u,x)\in\widetilde{X}_{k'}}\widetilde{\lambda}_{k'|k}(u,x)\label{eq:PPP_augmented}\\
\widetilde{\lambda}_{k'|k}\left(u,x\right) & =\delta_{0}\left[u\right]\lambda_{k'|k}\left(x\right)\\
\widetilde{f}_{k'|k}^{i,a^{i}}\left(\widetilde{X}_{k'}\right) & =\begin{cases}
1-r_{k'|k}^{i,a^{i}} & \widetilde{X}=\emptyset\\
r_{k'|k}^{i,a^{i}}p_{k'|k}^{i,a^{i}}\left(x\right)\delta_{i}\left[u\right] & \widetilde{X}=\left\{ \left(u,x\right)\right\} \\
0 & \mathrm{otherwise}\quad
\end{cases}\label{eq:Bernoulli_density_filter-aux}
\end{align}
where the Kronecker delta $\delta_{i}\left[u\right]=1$ if $u=i$
and $\delta_{i}\left[u\right]=0$, otherwise. The introduction of
the auxiliary variables allows us to remove the sum over the sets
in (\ref{eq:PMBM_aux_var2}), as there is only one term in the sum
that provides a non-zero density.
\end{defn}

\section{Clustered PMBM approximation via KLD minimisation\label{sec:Clustering_KLD}}

In this section, we are given clusters of potential targets, and our
aim is to approximate a PMBM density as a clustered PMBM density based
on KLD minimisation with auxiliary variables. The clustered PMBM density
ensures that potential targets belonging to different clusters are
independent, and corresponds to the union of an independent PPP and
independent MBMs, one for each cluster. The results in this section
hold for any clustering algorithm and show how to obtain the parameters
of the clustered PMBM posterior once the clusters are defined. The
specific data-driven clustering algorithm we propose is explained
in Section \ref{sec:Clustering_Measurements}.

\subsection{Clustered density\label{subsec:Clustered-density}}

Suppose we perform clustering at each update and denote the set of
clusters as $\mathcal{C}_{k}=\{C_{k}^{1},\dots,C_{k}^{n_{k}^{c}}\}$,
where each element $C_{k}^{i}$ is the set of auxiliary variables
corresponding to the tracks assigned to the cluster $c\in\{1,\dots,n_{k}^{c}\}$.
The set $\mathcal{C}_{k}$ is a partition of the auxiliary variable
space without ${0}$, $\mathbb{\mathbb{U}}_{k'|k}\setminus\left\{ 0\right\} $,
that meets the following properties \cite{Mahler14}
\begin{itemize}
\item Each cluster $C_{k}^{c}$ is a subset of auxiliary variables $C_{k}^{c}\subset\mathbb{\mathbb{U}}_{k'|k}\setminus\left\{ 0\right\} $.
\item The union of the clusters is the auxiliary variable space without
${0}$; i.e., $\cup_{c=1}^{n_{k}^{c}}C_{k}^{c}=\mathbb{\mathbb{U}}_{k'|k}\setminus\left\{ 0\right\} $.
\item The intersection of any two distinct clusters with indices $c_{1}$
and $c_{2}$, $c_{1}\neq c_{2}$, is empty; i.e., $C_{k}^{c_{1}}\cap C_{k}^{c_{2}}=\emptyset$.
\end{itemize}
Given $\mathcal{C}_{k}$, we can approximate the density of the set
of targets with independent clusters (including auxiliary variables)
as
\begin{align}
\widetilde{q}_{k'|k}\left(\widetilde{X}_{k'}\right) & =\widetilde{q}_{k'|k}^{0}\left(\widetilde{Y}_{k'}\right)\prod_{c=1}^{n_{k}^{c}}\widetilde{q}_{k'|k}^{c}\left(\cup_{i\in C_{k}^{c}}\widetilde{X}_{k'}^{i}\right).\label{eq:q_no_details_MBM}
\end{align}
where $\widetilde{q}_{k'|k}^{0}(\cdot)$ represents the density on
the set of undetected targets $\widetilde{Y}_{k'}$, and the density
on the detected target set is expressed as the multiplication of the
cluster densities $\widetilde{q}_{k'|k}^{c}(\cdot)$. The form (\ref{eq:q_no_details_MBM})
implies that targets belonging to different clusters are independent.
If there is exactly one potential target in each cluster, all potential
targets are independent, and (\ref{eq:q_no_details_MBM}) becomes
PMB, with auxiliary variables.

\subsection{Clustered PMBM with auxiliary variables\label{subsec:Clustered-PMBM-with}}

We calculate the clustered PMBM density (\ref{eq:q_no_details_MBM})
by minimising the KLD between $\widetilde{f}_{k'|k}\left(\cdot\right)$
in (\ref{eq:PMBM_aux_var2}) and $\widetilde{q}_{k'|k}\left(\cdot\right)$
in (\ref{eq:q_no_details_MBM}). The KLD is defined as the set integral
\cite{Mahler14}
\begin{align}
\mathrm{D}\left(\widetilde{f}_{k'|k}\left\Vert \widetilde{q}_{k'|k}\right.\right) & =\int\widetilde{f}\left(\widetilde{X}_{k'|k}\right)\log\frac{\widetilde{f}\left(\widetilde{X}_{k'|k}\right)}{\widetilde{q}\left(\widetilde{X}_{k'|k}\right)}\delta\widetilde{X}.\label{eq:KLD_f_q}
\end{align}

\begin{lem}
\label{lem:KLD_min}Let $\widetilde{f}_{k'|k}(\cdot)$ be the PMBM
density with auxiliary variables in (\ref{eq:PMBM_aux_var2}). The
densities $\widetilde{q}_{k'|k}^{0}\left(\cdot\right)$, $\widetilde{q}_{k'|k}^{1}\left(\cdot\right)$,...,
$\widetilde{q}_{k'|k}^{c_{k'|k}}\left(\cdot\right)$ in (\ref{eq:q_no_details_MBM})
that minimise the KLD $\mathrm{D}\left(\widetilde{f}_{k'|k}\left\Vert \widetilde{q}_{k'|k}\right.\right)$
are
\begin{align}
\widetilde{q}_{k'|k}^{0}\left(\widetilde{Y}_{k'}\right) & =\widetilde{f}_{k'|k}^{p}\left(\widetilde{Y}_{k'}\right)\label{eq:PoissonQ}\\
\widetilde{q}_{k'|k}^{c}\left(\cup_{i\in C_{k}^{c}}\widetilde{X}_{k'}^{i}\right) & \propto\sum_{a\in\mathcal{A}_{k'|k}}w_{k'|k}^{a}\prod_{i\in C_{k}^{c}}\left[\widetilde{f}_{k'|k}^{i,a^{i}}\left(\widetilde{X}_{k'}^{i}\right)\right].\label{eq:Q_MBM_prop}
\end{align}
where $\widetilde{Y}_{k'}$ and $\widetilde{X}_{k'}$ are given in
Definition \ref{def:aux_density}, and $\widetilde{q}_{k'|k}^{c}(\cdot)$
is the cluster density of the cluster $c$.
\end{lem}
Lemma 2 builds on (\ref{eq:PMBM_density}), which was derived in \cite{Williams15a},
and computes the clustered PMBM given the clusters. See App. \ref{sec:Proof-of-Lemma2}
for the proof of Lemma \ref{lem:KLD_min}. We can see that the density
of each cluster is a multi-Bernoulli mixture for the set of targets
in the cluster, and the density for the undetected targets is a PPP.
Therefore, (\ref{eq:q_no_details_MBM}) with (\ref{eq:PoissonQ})
and (\ref{eq:Q_MBM_prop}) define a clustered PMBM, with auxiliary
variables.

As in (\ref{eq:Q_MBM_prop}) index $i$ only goes through the potential
targets in the cluster, there can be repeated terms in the sum, which
can be merged into one. To do so, we can define a cluster alphabet
$\mathcal{A}_{k'|k}^{c}$ by only considering the entries of the $\mathcal{A}_{k'|k}$
that correspond to this cluster, adding a level of indirection between
the cluster density and the Bernoulli components that constitute them.
Then, we can define a weight for the $a_{c}$ cluster hypothesis that
is the sum over all the weights $w_{k'|k}^{a_{c}}$ with the same
local hypotheses for the potential targets in the cluster. Thus, we
can rewrite the cluster density (\ref{eq:Q_MBM_prop}) as
\begin{equation}
\widetilde{q}_{k'|k}^{c}\left(\cup_{i\in C_{k}^{c}}\widetilde{X}_{k'}^{i}\right)=\sum_{a_{c}\in\mathcal{A}_{k'|k}^{c}}w_{k'|k}^{a_{c}}\prod_{i\in C_{k}^{c}}\left[\widetilde{f}_{k'|k}^{i,a_{c}^{i}}\left(\widetilde{X}_{k'}^{i}\right)\right].\label{eq:Q_MBM}
\end{equation}

\subsection{Clustered PMBM\label{subsec:Clustered-PMBM}}

The clustered density without auxiliary variables can be obtained
by integrating out the auxiliary variables in (\ref{eq:q_no_details_MBM}).
\begin{lem}
\label{lem:IntOutAuxVar}Let $\widetilde{q}_{k'|k}\left(\cdot\right)$
be the clustered density with auxiliary variables in (\ref{eq:q_no_details_MBM})
defined in $\mathcal{F}\left(\mathbb{\mathbb{U}}_{k'|k}\times\mathbb{R}^{n_{x}}\right)$
. The corresponding density $q_{k'|k}\left(\cdot\right)$ in $\mathcal{F}\left(\mathbb{R}^{n_{x}}\right)$
is derived by integrating out the auxiliary variables, obtaining
\begin{align}
 & \sum_{u_{1:n}\in\mathbb{\mathbb{U}}_{k}^{n}}\widetilde{q}_{k'|k}\left(\left\{ \left(u_{1},x_{1}\right),...,\left(u_{n},x_{n}\right)\right\} \right)\nonumber \\
 & \quad=q_{k'|k}\left(\left\{ x_{1},...,x_{n}\right\} \right)\label{eq:integating_out_variables}
\end{align}
where
\begin{align}
q_{k'|k}\left(X_{k'}\right) & =\sum_{Y^{0}\uplus X^{1}\uplus...\uplus X^{c_{k'|k}}=X_{k'}}q_{k'|k}^{0}\left(Y^{0}\right)\prod_{c=1}^{n_{k}^{c}}q_{k'|k}^{c}\left(X^{c}\right)\label{eq:q_def}
\end{align}
and
\begin{align}
 & q_{k'|k}^{c}\left(\left\{ x_{1},...,x_{n}\right\} \right)\nonumber \\
 & \quad=\sum_{u_{1:n}\in\mathbb{\mathbb{U}}_{k}^{n}}\widetilde{q}_{k'|k}^{c}\left(\left\{ \left(u_{1},x_{1}\right),...,\left(u_{n},x_{n}\right)\right\} \right).\label{eq:q_c_def}
\end{align}
 
\end{lem}
The proof of Lemma \ref{lem:IntOutAuxVar} in reported in App. \ref{sec:Proof-of-Lemma3}.
If $q_{k'|k}^{0}\left(\cdot\right)$ and $q_{k'|k}^{c}\left(\cdot\right)$
are obtained via the KLD minimisation on a PMBM in (\ref{eq:PoissonQ})-(\ref{eq:Q_MBM_prop}),
the density $q_{k'|k}\left(\cdot\right)$ is the union of $c_{k'|k}+1$
independent RFS \cite{Mahler14} (as its density is obtained through
the convolution formula). One RFS represents undetected targets, and
each of the rest of them corresponds to the RFS in a cluster, whose
density is an MBM. Therefore, a clustered PMBM is the union of an
independent PPP and $c_{k'|k}$ independent MBMs.

It can be shown that the KLD between the PMBM (\ref{eq:PMBM_aux_var2})
and clustered PMBM (\ref{eq:q_no_details_MBM}) with auxiliary variables
is an upper bound of the KLD distance between the PMBM and clustered
PMBM densities without auxiliary variables \cite{GarciaFernandez2020a}
\begin{equation}
\mathrm{D}\left(f_{k'|k}\left\Vert q_{k'|k}\right.\right)\leq\mathrm{D}\left(\widetilde{f}_{k'|k}\left\Vert \widetilde{q}_{k'|k}\right.\right)\label{eq:KLD_f_q_aux}
\end{equation}
where $q_{k'|k}$ denotes the clustered PMBM density in the form of
(\ref{eq:q_def}) without auxiliary variables. Therefore, Lemma \ref{lem:KLD_min}
minimises an upper bound of the KLD between $f_{k'|k}$ and $q_{k'|k}$,
which is the one of primary interest.

\subsection{Recursive clustered PMBM approximation\label{subsec:Recursive-clustered-PMBM}}

So far, we have explained how to obtain a clustered PMBM density from
a PMBM density. In order to apply these results to the filtering recursion,
in this section we explain how to obtain a clustered PMBM from a previously
clustered PMBM, in which the clusters may differ.

After the prediction, we obtain a clustered PMBM density $\widetilde{f}_{k|k-1}$
of the form (\ref{eq:q_no_details_MBM}), where $\widetilde{q}_{k|k-1}^{0}$
and $\widetilde{q}_{k|k-1}^{c}$ are defined, respectively, in (\ref{eq:PoissonQ})
(\ref{eq:Q_MBM}) on the set of clusters $\mathcal{C}_{k-1}$. At
time $k$ we use a new cluster $\mathcal{C}_{k}$ (e.g., following
the procedure that will be described in Section \ref{subsec:Clustering}).
We compute a new clustered PMBM density $\widetilde{q}_{k|k-1}^{c'}$
based on the new set of clusters $\mathcal{C}_{k}$ via KLD minimisation,
where $c'$ is the cluster index in the set $\mathcal{C}_{k}$.
\begin{lem}
\label{lem:Recursive_clustered}Let us assume the predicted density
with auxiliary variables $\widetilde{f}_{k|k-1}\left(\cdot\right)$
is a clustered PMBM density with clusters $C_{k-1}^{1},...,C_{k-1}^{n_{k-1}^{c}}$
such that
\begin{align}
\widetilde{f}_{k|k-1}\left(\widetilde{X}_{k}\right) & =\widetilde{f}_{k|k-1}^{0}\left(\widetilde{Y}_{k}\right)\prod_{c=1}^{n_{k-1}^{c}}\widetilde{f}_{k|k-1}^{c}\left(\cup_{i\in C_{k-1}^{c}}\widetilde{X}_{k}^{i}\right)\label{eq:pred_clusteredPMBM}
\end{align}
where 
\begin{align}
\widetilde{f}_{k|k-1}^{c}\left(\cup_{i\in C_{k-1}^{c}}\widetilde{X}_{k}^{i}\right) & =\sum_{a_{c}\in\mathcal{A}_{k|k-1}^{c}}w_{k|k-1}^{a_{c}}\prod_{i\in C_{k-1}^{c}}\widetilde{f}_{k|k-1}^{i,a^{i}}\left(\widetilde{X}_{k}^{i}\right).\label{eq:pred_clusteredPMBM_component}
\end{align}
If the clusters at time step $k$ are $C_{k}^{1},...,C_{k}^{n_{k}^{c}}$,
the predicted clustered density $\widetilde{q}_{k|k-1}\left(\widetilde{X}_{k}\right)$
of the form (\ref{eq:q_no_details_MBM}) that minimises $D\left(\widetilde{f}_{k|k-1}||\widetilde{q}_{k|k-1}\right)$
is a clustered PMBM characterised by

\begin{align}
\widetilde{q}_{k|k-1}^{c'}\left(\cup_{i\in C_{k}^{c'}}\widetilde{X}_{k}\right) & \propto\prod_{c=1:C_{k}^{c'}\cap C_{k-1}^{c}\neq\emptyset}^{n_{k-1}^{c}}\sum_{a_{c}\in\mathcal{A}_{k|k-1}^{c}}w_{k|k-1}^{a_{c}}\nonumber \\
 & \quad\times\prod_{i\in C_{k}^{c'}\cap C_{k-1}^{c}}\widetilde{f}_{k|k-1}^{i,a^{i}}\left(\widetilde{X}_{k}^{i}\right).\label{eq:QMBM_update_prop}
\end{align}
and $\widetilde{q}_{k|k-1}^{0}$ as defined in (\ref{eq:PoissonQ}).
\end{lem}
See App. \ref{sec:Proof-of-Lemma4} for the proof of Lemma \ref{lem:Recursive_clustered}.
In (\ref{eq:QMBM_update_prop}), potential targets that belong to
the same cluster at time step $k-1$ and $k$ retain their statistical
dependencies (modelled by an MBM). The PMBM update is performed independently
for each cluster $c'$ on the basis of the predicted cluster density
(\ref{eq:QMBM_update_prop}) as described in \cite{Williams15a,Garcia-Fernandez18}.

\section{Measurement-driven clustering\label{sec:Clustering_Measurements}}

In this section we describe a novel procedure to efficiently cluster
the potential targets on the basis of the current set of measurements
$Z_{k}$. At each time step $k$, tracks and measurements are linked
through the gating procedure based on efficient data structures, described
in Section \ref{subsec:Gating}. The cluster formation is performed
by the algorithm described in Section \ref{subsec:Clustering}. Finally,
an efficient method to prune the global hypotheses in the new clustered
PMBM density is presented in Section \ref{subsec:Efficient-pruning}.

\subsection{Gating via efficient data structures\label{subsec:Gating}}

Gating can significantly reduce the complexity of the data association
problem by avoiding computing low-weight hypotheses \cite{Challa2009}.
In the PMBM update, each measurement can be associated to a previous
Bernoulli or to the PPP. For each previous Bernoulli, we calculate
its predicted measurement $\widehat{z}$ and its  covariance matrix
$S$ \cite{Garcia-Fernandez18}. In ellipsoidal gating, we can evaluate
if a received measurement $z_{j}$ is likely to be produced by the
hypothesis $a^{i}$ of the potential target $i$, $(i,a^{i})$, by
computing its Mahalanobis distance with the covariance matrix $S$,
to each $\widehat{z}$. For each hypothesis $(i,a^{i})$, $z_{j}\in Z_{k}$
belongs to $\mathcal{G}_{k}(i,a^{i})$ if the Mahalanobis distance
between $z_{j}$ and the predicted measurement of $(i,a^{i})$ is
below the threshold $\gamma_{G}$.

Denoting as $N_{k}^{h}$ the sum of the number of Bernoulli components
and the number of Gaussian components in the PPP intensity \cite{Garcia-Fernandez18}
in the filter at time instant $k$, the evaluation of all these possible
pairs has a complexity $\mathcal{O}(|Z_{k}|\cdot N_{k}^{h})$. It
is possible to lower the complexity of this process by using efficient
data structures; we proceed to discuss how $k$-d trees \cite{Bentley1975}
and R-trees \cite{Guttman1984} can be used in this context.

\subsubsection{$k$-d tree\label{subsec:-d-tree}}

The $k$-d tree is a binary space-partitioning tree, which recursively
divides the $k$-dimensional space to organise the entries and perform
fast range searches. At each time step $k$, the computational cost
of building a $k$-d tree based on the set of measurements $Z_{k}$
is $\mathcal{O}(|Z_{k}|\log|Z_{k}|)$. For each hypothesis $(i,a^{i})$,
we define the mean variance across dimensions in the innovation covariance
$(\sigma_{k}^{i,a^{i}})^{2}=\mathrm{tr}(S_{k}^{i,a^{i}})/n_{z}$.
Then, we perform $N_{k}^{h}$ range queries on the expected target
measurements $\widehat{z}_{k}^{i,a^{i}}$ for a range defined by $\gamma_{G}\sigma_{k}^{i,a^{i}}$.
Thus, the gating procedure queries the $k$-d tree in a computational
time $\mathcal{O}(N_{k}^{h}(|Z_{k}|^{1-1/n_{z}}+s))$ \cite{Uhlmann1992}
at each time instant, where $n_{z}$ is the number of dimensions of
the search space, and $s$ is the average number of measurements returned
by each query. Note that in our setting $k=n_{z}$ as the $k$-d tree
operates on the single measurement space.

\subsubsection{R-tree\label{subsec:R-tree}}

The R-tree is a hierarchical data structure in which every entry is
represented by a minimum bounding $d$-dimensional rectangle (MBR).
The internal nodes of the tree organise the leaf nodes into larger
MBR, allowing efficient retrievement of the entries that intersect
a window (or a point) in the $d$-dimensional space \cite{Manolopoulos2006}.

In the R-Tree implementation, the tree is built on the $N_{k}^{h}$
predicted single target states, with an overall computational time
$\mathcal{O}(N_{k}^{h}\log N_{k}^{h})$. The predicted measurement
from the Bernoulli $f^{i,a^{i}}$, and its covariance matrix are represented
by an $n_{z}$-dimensional rectangle $\mathcal{R}^{i,a^{i}}$ with
centre in $\widehat{z}_{k}^{i,a^{i}}$ and dimensions proportional
to the standard deviation $\sigma_{k}^{d,i,a^{i}}$ of the innovation
covariance $S_{k}^{i,a^{i}}$for each axis. That is, the gating area
is defined by
\begin{equation}
\mathcal{R}^{i,a^{i}}=\left\{ z:\left|z_{k}^{d}-\widehat{z}_{k}^{d,i,a^{i}}\right|\leq\gamma_{G}\sigma_{k}^{d,i,a^{i}},\forall d\right\} \label{eq:rect_box}
\end{equation}
where $d\in\{1,\dots,n_{z}\}$ indicates the dimensions, $z_{k}=[z_{k}^{1},...,z_{z_{k}}^{n}]^{T}$,
and $\gamma_{G}$ is the gating threshold. We define the set of measurements
$\mathcal{G}_{k}(i,a^{i})$ selected to update the hypothesis $a^{i}\in\{1,\dots,h_{k'|k}\}$
as the subset of the measurements $Z_{k}$ which belong to into the
rectangle $\mathcal{R}^{i,a^{i}}$. The gating procedure can be implemented
by inserting the hyper-rectangles $\mathcal{R}^{i,a^{i}}$ in the
R-tree, and it provides an efficient approximation of the ellipsoidal
gating \cite{Kurien90}. The entire gating procedure is performed
with $|Z_{k}|$ queries in a computational time $\mathcal{O}(|Z_{k}|((\log N_{k}^{h})^{n_{z}-1}+s))$,
where $s$ is the number of elements returned at each query. 

The capability of the R-tree to efficiently store hyper-rectangles
allows us to perform fewer queries than with the $k$-d tree, exploiting
the efficiency provided by the logarithmic query time on a greater
number of elements in the tree \cite{Uhlmann90}. On the other side,
the efficiency of the R-tree is reduced if the hyper-rectangles in
the structure show a high degree of overlap, as more edges need to
be inspected to complete a query \cite{Alborzi2007}.

\subsection{Clustering\label{subsec:Clustering}}

Potential targets that have local hypotheses with common measurements
at the current or past time steps are not independent and, in principle,
should belong to the same cluster. Nevertheless, the dependencies
in the distributions of the potential targets tend to weaken if there
are no common measurements in recent time steps. In this work, we
propose a clustering algorithm that only accounts for the data associations
at the current time step. The algorithm does not explicitly maintain
cluster information from scan to scan, as it defines new clusters
at each time step. It only retrieves information from the previous
time step to cluster misdetected tracks. Its main benefit is the computational
efficiency and ease of implementation, though it discards possible
target dependencies lingering from past time steps.

Suppose $\mathcal{C}_{k}=\{C_{k}^{1},\dots,C_{k}^{n_{k}^{c}}\}$ is
a partition of the auxiliary variable set $\mathbb{\mathbb{U}}_{k'|k}\backslash\{0\}=\left\{ 1,..,n_{k'|k}\right\} $.
For each cluster $C_{k}^{c}$ we can determine the set of associated
measurements $\mathcal{S}_{k}^{c}$ as the union of the sets of the
gated measurements for the targets in the cluster,
\begin{align}
\mathcal{S}_{k}^{c} & =\bigcup_{i\in C_{k}^{c}}\mathcal{G}_{k}^{i}\label{eq:union_of_meas}
\end{align}
where $\mathcal{G}_{k}^{i}=\left\{ \mathcal{G}^{k}(i,a^{i})|a^{i}\in\{1,\dots,h_{k'|k}^{i}\}\right\} $
is the set of measurements related to the track $i$. 

\begin{figure}
\centering{}\includegraphics[scale=0.3]{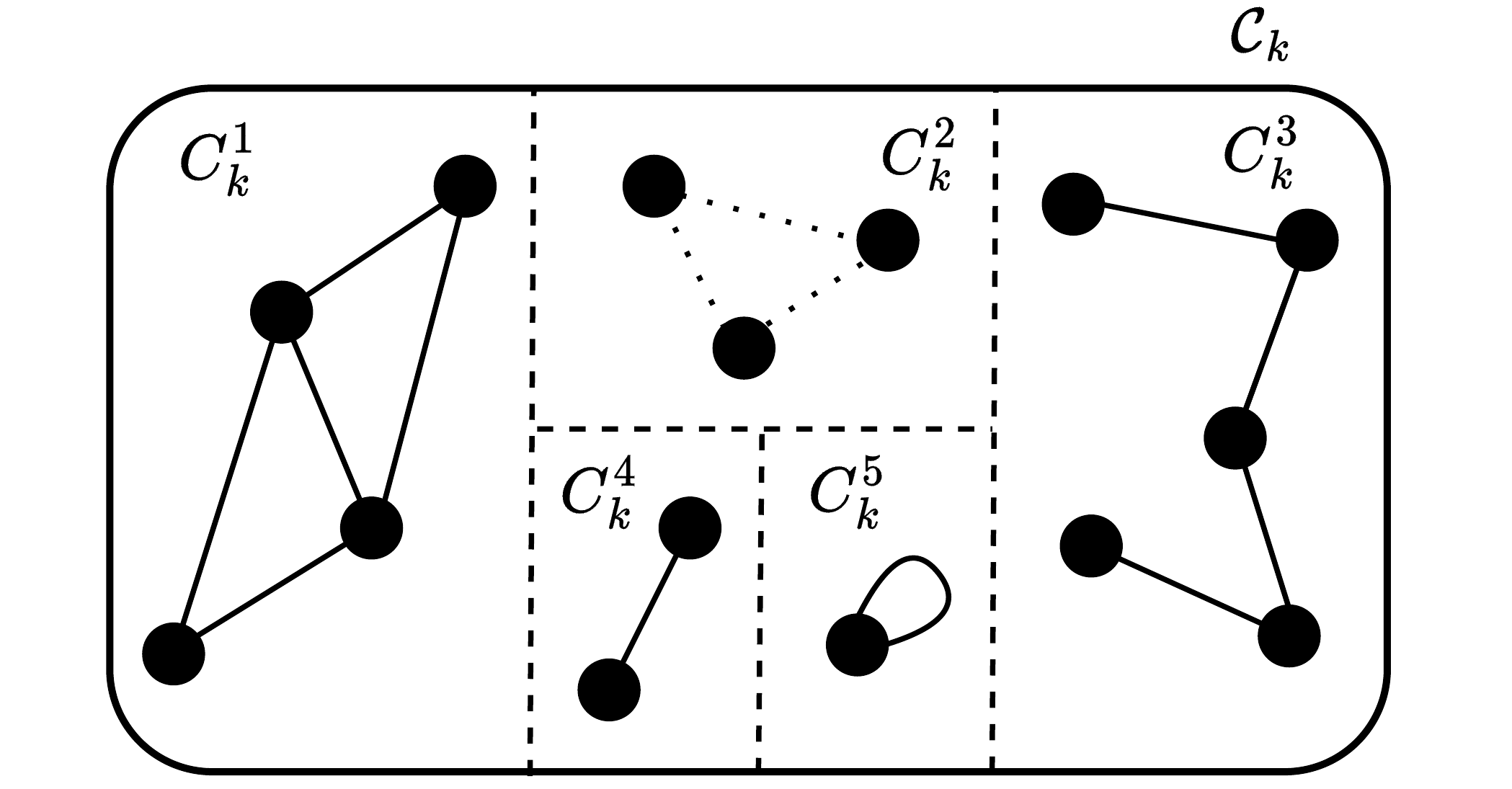}\caption{Example of disjoint union of graphs. The 15 nodes represent the potential
targets arranged in 5 clusters, on the basis of the common measurements
represented by the edges. Adjacent nodes depict potential targets
associated to the same measurement, while loops represent measurements
related to a single potential target. Cluster $C_{k}^{2}$ contains
three misdetected targets, which belonged to the same cluster at the
previous time instant. The dashed lines in $C_{k}^{2}$ represent
the dummy measurement assigned by the clustering algorithm (Alg. \ref{alg:Measurement-driven-clustering}).
\label{fig:Example-of-disjoint}}
\end{figure}

The relation between targets and measurements can be represented by
a graph, where the nodes denote the targets, and the edges connect
targets that have at least one measurement in their gates in common.
The partitioning of the tracks into clusters is defined by the connected
components of the graph, which can be considered as a disjoint union
of graphs (see Fig. \ref{fig:Example-of-disjoint}). The connected
components of the graph can be determined by an algorithm for traversing
or searching graph data structures, as depth-first search or breadth-first
search \cite{MichaelT.Goodrich2001}. This approach obtains sets of
clusters such that the intersection of any two distinct measurements
sets is empty; i.e., $\mathcal{S}_{k'|k}^{c_{1}}\cap\mathcal{S}_{k'|k}^{c_{2}}=\emptyset$,
$c_{1}\neq c_{2}$, $\{c_{1},c_{2}\}\subset\{1,\dots n_{k}^{c}\}$.

The isolated nodes of the graph, i.e., those that are not an endpoint
of any edge, represent the misdetected targets at the current time
step $k$, as they are not associated with any measurement. The algorithm
clusters the misdetected targets according to their cluster membership
at the previous time instant $k-1$.  A dummy measurement $\ast_{j}$,
$j\in\{1,\dots n_{k-1}^{c}\}$, is generated for each cluster at time
$k-1$, and the targets misdetected at time $k$ are associated with
the corresponding measurement $\ast_{j}$, where $j$ represent the
cluster membership at $k-1$. Fig. \ref{fig:Example-of-disjoint}
shows an example of this kind for cluster $C_{k}^{2}$, where it can
be noted that the misdetected targets are clustered due to the links
provided by the dummy measurement, represented with dashed lines.
The pseudocode of the clustering algorithm is provided in Algorithm
\ref{alg:Measurement-driven-clustering}.

\begin{algorithm}
\textbf{Input: }Pairs target-measurements $A=\{(i,\mathcal{G}_{k}^{i})|i\in\mathbb{\mathbb{U}}_{k'|k}\}$\\
\hspace*{\algorithmicindent}\hspace*{\algorithmicindent}Set of clusters at the previous time step \\
\hspace*{\algorithmicindent}\hspace*{\algorithmicindent}$\mathcal{C}_{k-1}=\{C^1,\dots,C^{n^c_{k-1}}\}$\\
\textbf{Output: }Set of clusters at the current time step $\mathcal{C}_{k}$
\begin{algorithmic}[1]
\State Obtain $U_{0}=\{i:i\in\mathbb{\mathbb{U}}_{k|k}\setminus\left\{0\right\},\mathcal{G}_{k}^{i}=\emptyset\}$: the set of indices of misdetected tracks.
\State Retrieve the indices of misdetected tracks at time step $k$ in each cluster defined at the previous time step $\{C_0^1,\dots,C_0^{n^c_{k-1}}\}$ such that $C_0^i\subseteq C^i$, $\cup_{i=1}^{n^c_{k-1}}C_0^{i}=U_{0}$.
\For{$j\in\{1\dots,n^c_{k-1}\}$}
\For{$i\in C_0^j$}
\State $\mathcal{G}^i_k \gets \mathcal{G}^i_k \cup \{\ast_j\}$ \Comment{Assign a dummy measurement.}
\EndFor
\EndFor
\State Generate the graph $G$ on $A$.
\State Assign the track indices of the connected components of $G$ to a cluster to obtain $\mathcal{C}_{k}=\{C^1,\dots,C^{n^c_k}\}$.\\
\Return Set of clusters $\mathcal{C}_{k}$.
\end{algorithmic}\caption{Measurement-driven clustering\label{alg:Measurement-driven-clustering}}
\end{algorithm}

\subsection{Efficient pruning for the new clustered PMBM\label{subsec:Efficient-pruning}}

After we have obtained the clusters via Algorithm \ref{alg:Measurement-driven-clustering},
we can use Lemma \ref{lem:KLD_min} or \ref{lem:Recursive_clustered}
to obtain the clustered PMBM, and perform the update for each cluster
independently. Due to the product over the clusters in (\ref{eq:QMBM_update_prop}),
if previously independent clusters are merged, the resulting MBM can
contain a high number of multi-Bernoulli components. We propose an
efficient method to prune the least likely components by computing
only the $K$ best merged global hypotheses in each cluster $C_{k}^{c'}\in\mathcal{C}_{k}$,
where $K$ is adaptively determined by the normalised merged weights. 

Suppose $c'$ is the index of the new cluster and assume we merge
the previous clusters with indices in the set $\mathcal{M}=\{c|C_{k}^{c'}\cap C_{k-1}^{c}\neq\emptyset\}$.
We define the binary decision variables $v_{c,j}\in\{0,1\}$, where
$v_{c,j}=1$ if the global hypothesis of index $j\in\{1,\dots,|\mathcal{A}_{k|k-1}^{c}|\}$
in the cluster $c$ contributes to the solution. According to the
same logic, we denote the weight selected by the variable $v_{c,j}$
as $w_{c,j}$. We can describe the problem of finding the best merged
global hypothesis as an optimisation problem 
\begin{align}
\text{maximise } & \sum_{c\in\mathcal{M}}\sum_{j=1}^{\left|\mathcal{A}_{k|k-1}^{c}\right|}v_{c,j}\log w_{c,j}\label{eq:opt_pruning}\\
\text{subject to } & \sum_{j=1}^{\left|\mathcal{A}_{k|k-1}^{c}\right|}v_{c,j}=1,\quad c\in\mathcal{M}\,.\label{eq:cond1}
\end{align}
The solution to (\ref{eq:opt_pruning}) corresponds to taking the
maximum weight $w_{c,j}$ over $j$ for each $c$. Here, we take the
$K$ best global hypothesis until the normalised weight of the $K$-th
hypothesis is below a pruning threshold $\Gamma_{mbm}$.

Assuming the set of global hypotheses is sorted by descending weight
in each cluster $c\in\mathcal{M}$, the problem can be solved in $\mathcal{O}(\left|\mathcal{M}\right|K\max(\log\left|\mathcal{M}\right|,\log K)$)
using a branch-and-bound approach implemented with a priority queue
\cite[Ch. 6]{Neapolitan2014}. Starting from the best hypothesis,
defined as the product of the best hypothesis in each cluster, we
determine the other $K-1$ merged hypotheses by iteratively extracting
and expanding the best solution in the tree of all the possible combinations
of weights.

Once we obtain the set of merged global hypotheses for each cluster
$c\in\mathcal{M}$, we can define the posterior density on a partition
$\mathcal{C}_{k},$ as shown in Lemma \ref{lem:KLD_min} and \ref{lem:Recursive_clustered}.
If the $n_{k'|k}$ tracks are partitioned into a set of $n_{k}^{c}$
clusters, the simplified data association problem can be split into
$n_{k}^{c}$ independent subproblems. This approximation enables a
remarkable reduction of the computational time and it enables the
direct use of parallelization techniques in the update step.

\section{Merging of Bernoulli densities\label{sec:Merging}}

In this section, we present two merging strategies to reduce the number
of Bernoulli components in the clustered PMBM posterior. In Section
\ref{subsec:Local-merging}, we propose to merge the most similar
single target hypotheses corresponding to the same potential target
according to the KLD between Bernoulli RFSs \cite{Fontana2020}. In
Section \ref{subsec:Trans-Bernoulli-merging}, we present an algorithm
that can rearrange the Bernoulli components across different potential
targets that is useful in cluster formation and to lower the number
of Bernoulli components in situations where targets get in close proximity
and then separate.

\subsection{Intra-track Bernoulli merging\label{subsec:Local-merging}}

This section deals with merging of different local hypotheses corresponding
to the same potential target. The aim is to detect the Bernoulli components
that are sufficiently similar in terms of KLD for each potential target.
The similar Bernoulli components, each in a different local hypothesis,
are substituted by a single local hypothesis, reducing the overall
number of single target hypotheses and decreasing the computational
burden of the update step. 

The algorithm results in an heuristic mixture reduction procedure
which iteratively decrease the number of mixture components by merging
the most similar Bernoulli components at each iteration. Note that,
unlike in \cite{Xia2022}, the merging is performed at the Bernoulli
components level, and it does not affect the global hypotheses weights
directly. Nevertheless, after the update of the global hypotheses
with the new local hypotheses indices due to merging, the list of
global hypotheses can present duplicates, which can be simplified
by summing the weights of the identical global hypotheses.

We perform merging in two ways. First, Bernoulli components (of the
same potential target) associated with the same measurement are merged.
Second, we use the KLD to determine similar Bernoulli components that
should be merged. Note that a distance between Bernoulli densities
based on the Rényi divergence has been presented in \cite{Ristic2012}.

\subsubsection{Bernoulli merging\label{subsec:Bernoulli-merging}}

Let us consider a potential target $i$, its $h_{k|k}^{i}$ single
target hypotheses with index $a^{i}\in\{1,\dots,h_{k|k}^{i}\}$, and
the subset of global hypotheses $\mathcal{\mathcal{A}}{}_{k'|k}^{i}\subseteq\mathcal{\mathcal{A}}{}_{k'|k}$
in which the target $i$ is supposed to exist. The potential target
state can be described by the mixture of Bernoulli densities
\begin{equation}
\widetilde{f}_{k'|k}^{i}(\widetilde{X}^{i})=\sum_{a\in\mathcal{\mathcal{A}}{}_{k'|k}^{i}}W_{k'|k}^{i,a^{i}}\widetilde{f}_{k'|k}^{i,a^{i}}(\widetilde{X}^{i})\label{eq:mixture}
\end{equation}
where $W_{k'|k}^{i,a^{i}}$ is the component weight defined as the
sum of the weights associated to the global hypotheses in which a
specific Bernoulli component $\widetilde{f}_{k|k}^{i,a^{i}}$ appears,
i.e.,
\begin{equation}
W_{k'|k}^{i,a^{i}}=\sum_{a\in\mathcal{\mathcal{A}}{}_{k'|k}^{i}}w_{k'|k}^{a}\text{ .}\label{eq:sumW}
\end{equation}
Suppose $p_{k'|k}^{i,a^{i}}(x)$, the single target density of $\widetilde{f}_{k'|k}^{i,a^{i}}(\widetilde{X}^{i}),$
is Gaussian, e.g. $p_{k'|k}^{i,a^{i}}(x)=\mathcal{N}(x;\mu_{k'|k}^{i,a^{i}},P_{k'|k}^{i,a^{i}})$.
Assume that we aim to merge the components of indexes $\mathcal{A}_{k'|k}^{i_{m}}\subseteq\{1,\dots,h_{k|k}^{i}\}$
in the Bernoulli mixture $\widetilde{f}_{k'|k}^{i}(\widetilde{X}^{i})$
into a single Bernoulli density $\widehat{f}_{k'|k}^{i}\left(\widetilde{X}^{i}\right)$,
with single target density $\widehat{p}_{k'|k}^{i}(x)=\mathcal{N}(x;\widehat{\mu}_{k'|k}^{i,},\widehat{P}_{k'|k}^{i})$.
The approximated Bernoulli density $\widehat{f}_{k'|k}^{i}\left(\widetilde{X}^{i}\right)$
that minimises the KLD $D(\widetilde{f}|\widehat{f}$) is characterised
by 
\begin{equation}
\widehat{W}_{k'|k}^{i,a^{i}}=\sum_{a^{i}\in\mathcal{A}_{k'|k}^{i_{m}}}W_{k'|k}^{i,a^{i}}\label{eq:merged_component_weight}
\end{equation}
and it is expressed by \cite{Williams15a,Xia2022}:
\begin{align}
\widehat{f}_{k'|k}^{i}\left(\widetilde{X}^{i}\right) & =\begin{cases}
1-\widehat{r}_{k'|k}^{i} & \widetilde{X}^{i}=\emptyset\\
\widehat{r}_{k'|k}^{i}\mathcal{N}(x;\widehat{\mu}_{k'|k}^{i},\widehat{P}_{k'|k}^{i})\delta_{i}\left[u\right] & \widetilde{X}^{i}=\left\{ \left(u,x\right)\right\} \\
0 & \mathrm{otherwise}
\end{cases}\label{eq:Bernoulli_density_filter_merged}
\end{align}

where
\begin{equation}
\widehat{r}_{k'|k}^{i}=\frac{\sum_{a^{i}\in\mathcal{A}_{k'|k}^{i_{m}}}W_{k'|k}^{i,a^{i}}r_{k'|k}^{i,a^{i}}}{\sum_{a^{i}\in\mathcal{A}_{k'|k}^{i_{m}}}W_{k'|k}^{i,a^{i}}}\label{eq:new_prob_ext}
\end{equation}
\begin{equation}
\mathbf{\widehat{\mu}}_{k'|k}^{i}=\frac{\sum_{a^{i}\in\mathcal{A}_{k'|k}^{i_{m}}}W_{k'|k}^{i,a^{i}}r_{k'|k}^{i,a^{i}}\mu_{k'|k}^{i,a^{i}}}{\sum_{a^{i}\in\mathcal{A}_{k'|k}^{i_{m}}}W_{k'|k}^{i,a^{i}}r_{k'|k}^{i,a^{i}}}\label{eq:new_mean}
\end{equation}
\begin{align}
\widehat{P}_{k'|k}^{i} & =\frac{\sum_{a^{i}\in\mathcal{A}_{k'|k}^{i_{m}}}W_{k'|k}^{i,a^{i}}r_{k'|k}^{i,a^{i}}(P_{k'|k}^{i,a^{i}}+\mu_{k'|k}^{i,a^{i}}(\mu_{k'|k}^{i,a^{i}})^{T})}{\sum_{a^{i}\in\mathcal{A}_{k'|k}^{i_{m}}}W_{k'|k}^{i,a^{i}}r_{k'|k}^{i,a^{i}}}\nonumber \\
 & \quad-\mathbf{\widehat{\mu}}_{k'|k}^{i}(\mathbf{\widehat{\mu}}_{k'|k}^{i})^{T}\text{ .}\label{eq:new_cov_matrix}
\end{align}

\subsubsection{KLD between Bernoulli distributions\label{subsec:KLD-between-Bernoulli}}

We aim to find an approximation of ($\ref{eq:mixture}$) by merging
the most similar Bernoulli components. We evaluate the similarity
between two Bernoulli distributions using the closed-form of the KLD
between two Bernoulli distributions presented in Lemma \ref{lem:KLD}.
The proof and other distances for Bernoulli merging are available
in \cite{Fontana2020}.
\begin{lem}
Let $\widetilde{f}_{1}(\widetilde{X})$ and $\widetilde{f}_{2}(\widetilde{X})$
be two Bernoulli RFS distributions with Gaussian single target densities.
The $i$-th Bernoulli RFS has probability of existence $r_{i}$, mean
$\bar{x}_{i}$, and covariance matrix $P_{i}$. If $r_{2}\notin\{0,1\}$,
the KLD of $\widetilde{f}_{2}$ from $\widetilde{f}_{1}$ exists and
it is a finite value, given by:\label{lem:KLD}
\begin{align}
\mathrm{D}_{KL} & (\widetilde{f}_{1}\left\Vert \widetilde{f}_{2}\right.)\nonumber \\
 & =\left(1-r_{1}\right)\log\frac{1-r_{1}}{1-r_{2}}+r_{1}\log\frac{r_{1}}{r_{2}}\nonumber \\
 & \quad+\frac{r^{1}}{2}\left[\mathrm{tr}\left(\left(P_{2}\right)^{-1}P_{1}\right)-\log\left(\frac{\left|P_{1}\right|}{\left|P_{2}\right|}\right)-n_{x}\right.\nonumber \\
 & \quad\left.+\left(\overline{x}_{2}-\overline{x}_{1}\right)^{T}\left(P_{2}\right)^{-1}\left(\overline{x}_{2}-\overline{x}_{1}\right)\vphantom{\left(\frac{\left|P^{2}\right|}{\left|P^{2}\right|}\right)}\right]\text{ .}\label{eq:KLD-1}
\end{align}

If $r_{1}=r_{2}\in\{0,1\}$, the KLD is:
\begin{align}
\mathrm{D}_{KL} & (\widetilde{f}_{1}\left\Vert \widetilde{f}_{2}\right.)\nonumber \\
 & =\frac{r^{1}}{2}\left[\mathrm{tr}\left(\left(P_{2}\right)^{-1}P_{1}\right)-\log\left(\frac{\left|P_{1}\right|}{\left|P_{2}\right|}\right)-n_{x}\right.\nonumber \\
 & \quad\left.+\left(\overline{x}_{2}-\overline{x}_{1}\right)^{T}\left(P_{2}\right)^{-1}\left(\overline{x}_{2}-\overline{x}_{1}\right)\vphantom{\left(\frac{\left|P^{2}\right|}{\left|P^{2}\right|}\right)}\right]\text{ .}\label{eq:KLD2}
\end{align}
\end{lem}

\subsubsection{Identification of similar Bernoulli components\label{subsec:Identification-of-similar}}

The proposed intra-track merging algorithm consists of two main steps.
Firstly, for each potential target $i$, the algorithm reduces the
$h_{k|k-1}^{i}$ Bernoulli components associated with $z_{k}^{j}$
to one single Bernoulli component $\widehat{f}_{k|k}^{i,j}$ by moment-matching,
see (\ref{eq:new_prob_ext})-(\ref{eq:new_cov_matrix}). The output
of this step is a set of $m_{k}$ single target hypotheses resulting
from the merging algorithm, $h_{k|k-1}^{i}$ Bernoulli components
associated with a misdetection hypothesis, and their relative weights
in the mixture.

Secondly, the single target hypotheses are iteratively merged by following
a greedy merge procedure based on the KLD defined in Lemma \ref{lem:KLD},
as described in \cite{Fontana2020}. The procedure considers the distances
between all the elements of the Bernoulli set, and merges the two
most similar Bernoulli components at each iteration, i.e., those which
show the minimum distance. The merging is performed only if the distance
is below a specified threshold $\Gamma_{m}$. Otherwise, the algorithm
breaks the loop and returns the current set of Bernoulli components.
The algorithm is similar to that proposed by Runnalls in \cite{Runnalls07},
and the pseudocode is available in \cite{Fontana2020}. Note that
an appropriate choice of the threshold $\Gamma_{m}$ allows the filter
to keep well-spaced mixture components in the mixture, providing a
more adequate representation of complex scenarios.

\begin{figure}
\centering{}\includegraphics[scale=0.5]{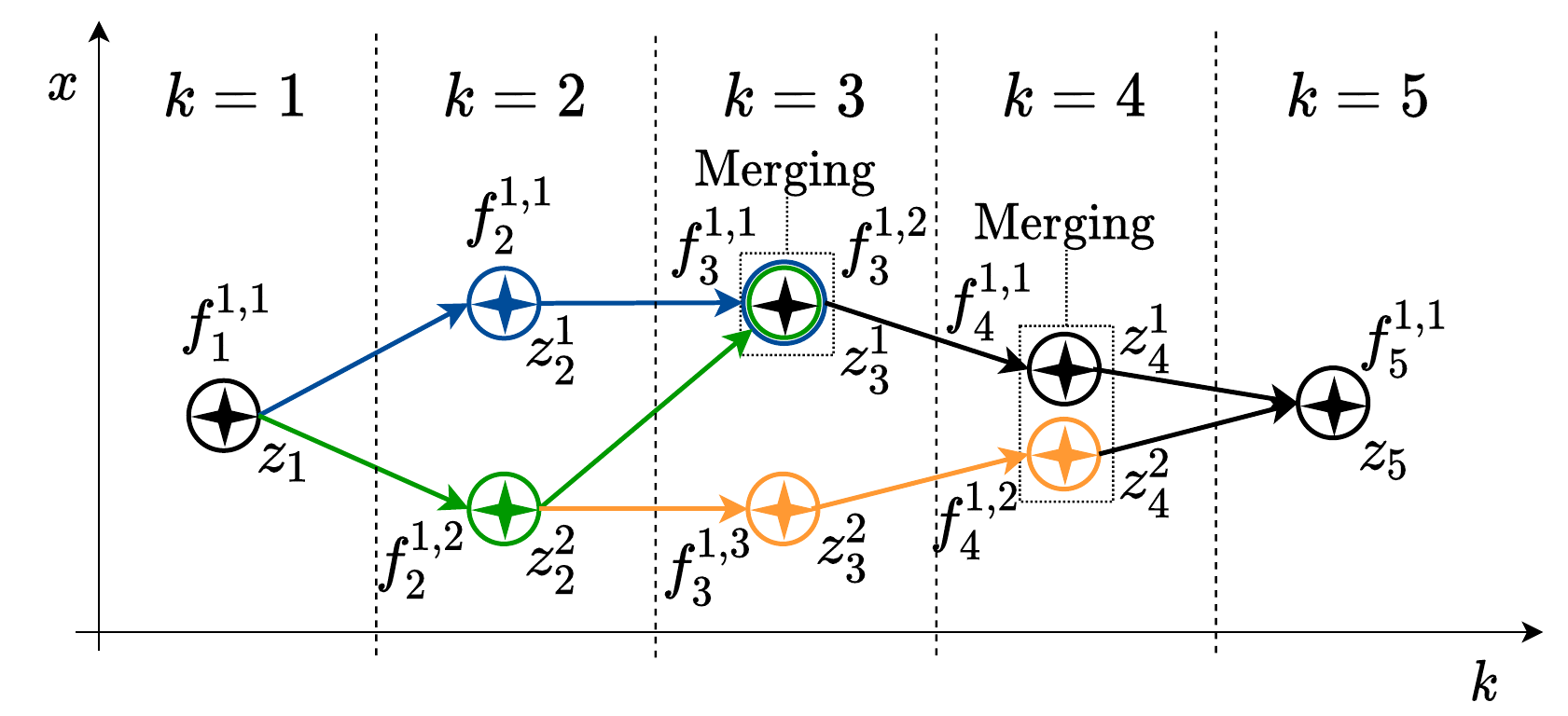}\caption{\label{fig:dev_diag}Example of a superposition of Bernoulli components
for the same potential target. The stars indicate the measurements
$z_{k}$ at time $k$, while the circles represent the Bernoulli components
associated to the corresponding measurement at each time step. At
$k=3$, $f_{3}^{1,1}$ and $f_{3}^{1,2}$ are updated with the same
measurement $z_{3}^{1}$, and they are merged into the new Bernoulli
$\widehat{f}_{3}^{1,1}$(not displayed in the figure). At the next
time instant, the KLD between $f_{4}^{1,1}$ and $f_{4}^{1,2}$ results
below the threshold, leading to the merging for the local hypothesis.
The notation for the density $\widetilde{f}(\cdot)$ has been simplified
in the figure.}
\end{figure}

Fig. \ref{fig:dev_diag} shows an example of the different steps of
the intra-track merging algorithm. At time $k=3$, the potential target
$i=1$ is described by three Bernoulli components, of which two updated
with the same measurement $z_{3}^{1}$. We can assume that both $f_{3}^{1,1}$
and $f_{3}^{1,2}$ are sufficiently similar, and merge them in a new
hypothesis according to the first step of the procedure. At time $k=4$,
the KLD between the two target hypotheses is lower than a pre-defined
threshold, enabling the reduction via moment-matching to one single
component.

\subsection{Inter-track Bernoulli swapping \label{subsec:Trans-Bernoulli-merging}}

It is known that after targets get in close proximity and then separate,
local targets hypotheses from different targets get mixed up. For
example, in an area where there is only one target, we may have hypotheses
from multiple targets \cite{Kreucher05}. A standard clustering algorithm
cannot put this target into a single cluster.

In this section, we propose a strategy to swap Bernoulli components
across different potential targets to improve clustering after targets
get in close proximity and separate. There are two computational benefits
of swapping Bernoulli components when targets get in close proximity
and the separate 1) we avoid performing repeated prediction and update
steps that are similar for different potential targets in standard
PMBM filtering, see Fig. \ref{fig:Example-target-crossing}, and 2)
we increase the number of clusters.

We first note that the PMBM posterior (without auxiliary variables)
remains unchanged by permuting the Bernoulli indices in each global
hypothesis. That is, the clustered MBM in (\ref{eq:Q_MBM}) expressed
without auxiliary variables is equivalent to
\begin{align}
\widehat{q}_{k'|k}^{c} & \left(\cup_{i\in C_{k}^{c}}X_{k'}^{i}\right)=\nonumber \\
 & \sum_{a_{c}\in\mathcal{A}_{k'|k}^{c}}w_{k'|k}^{a_{c}}\sum_{\cup_{j\in C_{k}^{c}}X^{j}=X}\prod_{i\in C_{k}^{c}}\left[f_{k'|k}^{\sigma_{a_{c}}(i),a_{c}^{\sigma_{a_{c}}(i)}}\left(X_{k'}^{i}\right)\right]\label{eq:Q_MBM_swap}
\end{align}
where $\sigma_{a_{c}}=(\sigma_{a_{c}}(1),...,\sigma_{a_{c}}(n_{k'|k}^{c}))$
is a permutation of $(1,...,n_{k'|k}^{c})$ applied to global hypothesis
$a_{c}$. The idea is then to use the flexibility introduced in (\ref{eq:Q_MBM_swap})
to design a fast algorithm that swaps the candidate Bernoulli indices
in specific global hypotheses to improve clustering. In addiction,
these candidates will then likely be merged by the intra-track Bernoulli
merging algorithm, described in Section \ref{subsec:Local-merging},
at the next time step.  In the following, we propose a computationally
efficient method to exploit this flexibility through four steps.

\subsubsection{Candidate tracks\label{subsec:Candidate-tracks}}

We identify the tracks with divergent data association histories indirectly
by computing the KLD between the Gaussian component of the single
target hypotheses of each track in a cluster. If a pair of components
of track $i$ has a KLD greater than a defined threshold $\Gamma_{s}$,
we consider that the data association history is divergent and the
track $i$ is considered as candidate for the inter-track swapping.
Given the Gaussian component $p_{k'|k}^{i,a^{i}}$ of the single target
hypothesis $a^{i}$ of track $i$, we define the set of candidate
tracks in the cluster $c\in\{1,\dots,n_{k}^{c}\}$ as
\begin{equation}
T_{c}=\left\{ i|\mathrm{D}_{KL}\left(p_{k'|k}^{i,a^{i}}\left\Vert p_{k'|k}^{i,b^{i}}\right.\right)>\Gamma_{s}\right\} \label{eq:candidate_tracks}
\end{equation}
where $i\in C_{k}^{c}$ and $a^{i},b^{i}\in\{1,\dots h_{k'|k}^{i}\}$.
Note that this procedure can be implemented as an extension of the
intra-track merging procedure presented in Section \ref{subsec:Local-merging}
with no extra computational time.

In Fig. \ref{fig:Example-target-crossing}, we consider $T_{c}=\{1,2\}$,
as we suppose that the KLD between the Bernoulli components of the
tracks 1 and 2 exceeds the threshold at time $k=k_{2}$.

\subsubsection{Bernoulli local clustering\label{subsec:Bernoulli-local-clustering}}

We seek to represent each potential target $i\in T_{c}$ by means
of a set of similar hypotheses, i.e., Bernoulli components located
in the same area. We apply the K-means algorithm \cite{Hastie09}
on the posterior mean positions of the candidate tracks to obtain
a partition of the Bernoulli components, where K$=|T_{c}|$. We indicate
the resulting local cluster associated with the Bernoulli component
$f_{k'|k}^{i,a_{c}^{i}}(\cdot)$ with the index $j^{i,a_{c}^{i}}\in\{1,\dots,|T_{c}|\}.$
The clustering requires a low increase in the computational burden
of the algorithm, as the number of local hypotheses to cluster is
usually low due to the pruning and merging procedures applied before
this point. The example in Fig. \ref{fig:Example-target-crossing}
shows the partition of the Bernoulli components in the cluster $C_{k_{2}}^{1}$
into two subclusters $G_{1}$ and $G_{2}$, where the subscripts $\{1,2\}$
represent the indices $j^{i,a_{c}^{i}}$.

\subsubsection{Track assignment to local clusters\label{subsec:Track-assignment-to}}

At this point, each candidate track belongs to multiple local clusters.
We assign each track $i\in T_{c}$ to one single local cluster, in
order to locate it in a specific area. We express the assignment by
the vector $s=\left(s(1),\dots,s(|T_{c}|)\right)$, $s(j^{i,a_{c}^{i}})\in T_{c}$,
with $s(j^{i,a_{c}^{i}})$ being the index of the reference track
for the local cluster $j^{i,a_{c}^{i}}$. This procedure allows us
to split the original cluster into several subclusters, correspondent
to the local clusters, reducing the data association problem into
smaller ones.

\subsubsection{Bernoulli swapping\label{subsec:Bernoulli-swapping}}

At this stage, we allocate the Bernoulli components in each local
cluster $l\in\{1,\dots,|T_{c}|\}$ to the track $s(l)$ assigned to
the local cluster $l$. 

\begin{figure}
\centering{}\includegraphics[scale=0.6]{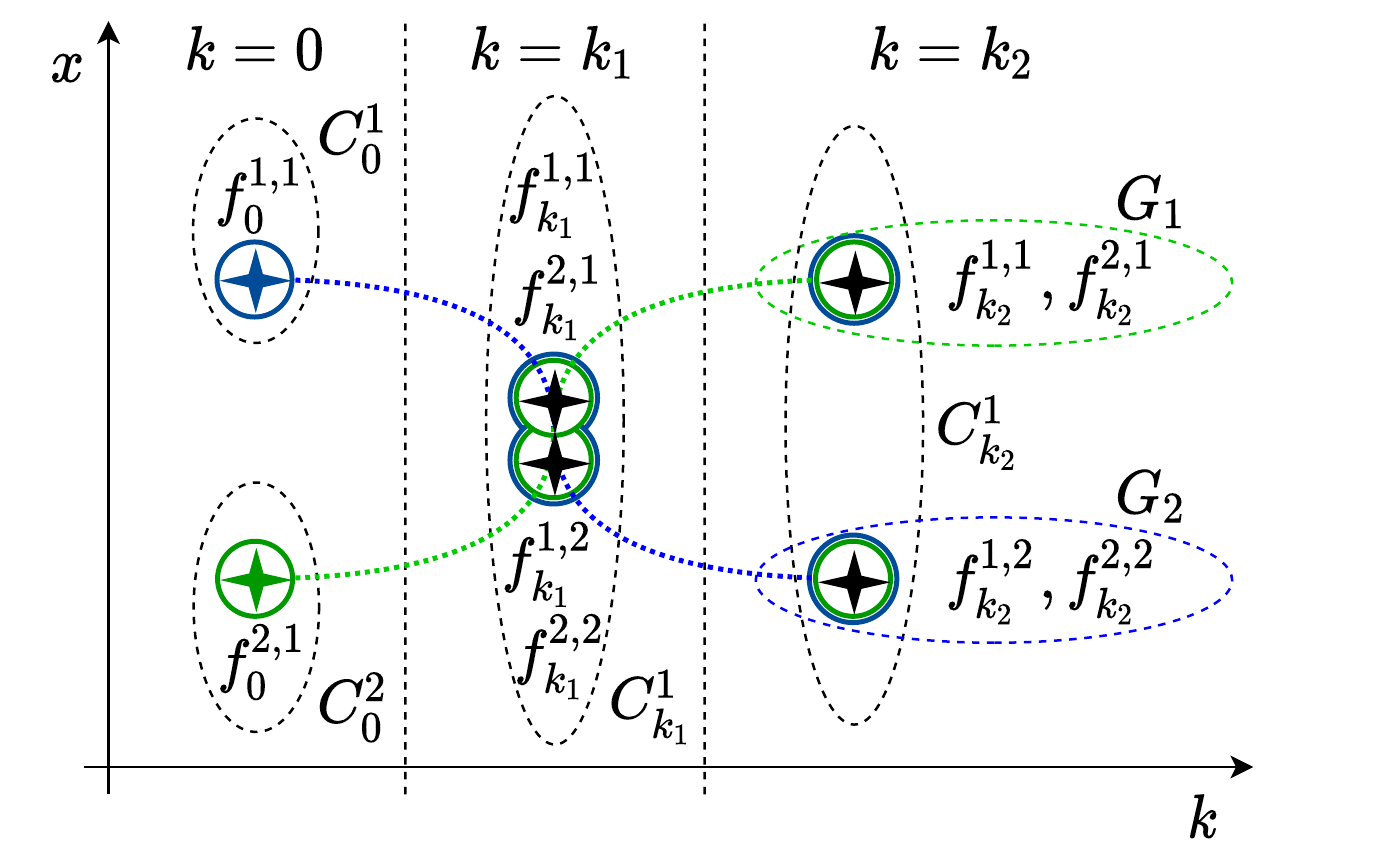}\caption{Example of two targets crossing. The stars indicate the measurements
at time $k=\{0,k_{1},k_{2}\}$, and the coloured circles represent
the Bernoulli components associated to the tracks at each time step,
where the tracks $i=\{1,2\}$ are depicted respectively in blue and
green. The tracks are composed of a single Bernoulli at $k=0$, and
then updated with common measurements at $k=k_{1}$. When the targets
move away, the tracks are represented at different locations, and
the related Bernoulli components can be clustered into two local clusters,
namely $G_{1}$ and $G_{2}$. The notation for the density $\widetilde{f}(\cdot)$
has been simplified in the figure. \label{fig:Example-target-crossing}}
\end{figure}

For example, assume the track $1$ is assigned to the local cluster
$G_{1}$ and track $2$ is assign to $G_{2}$ in Fig. \ref{fig:Example-target-crossing}.
The swapping procedure aims to allocate $f_{k_{2}}^{2,1}$ to track
1 and $f_{k_{2}}^{1,2}$ to track 2.

We represent the Bernoulli swapping procedure by considering the MBM
cluster density of the current cluster $c$ expressed in (\ref{eq:Q_MBM}).
We  determine an equivalent MBM cluster density as in (\ref{eq:Q_MBM_swap})
by defining $\sigma_{a_{c}}$ according to the following rules:
\begin{itemize}
\item $\sigma_{a_{c}}(i)=i$, for $i\notin T_{c}$.
\item $\sigma_{a_{c}}(i)=s(j^{i,a_{c}^{i}})$ for $i\in T_{c}$.
\end{itemize}

Once we have selected $\sigma_{a_{c}}$, we should note that there
is a rearrangement between Bernoulli components and tracks that carries
along to the following time steps.

\section{Simulations\label{sec:Simulations}}

In this section, we proceed to assess the accuracy and computational
time of the clustered PMBM filter and the proposed Bernoulli merging
strategies in two scenarios. We also compare the standard PMBM filter
implementation  \cite{Garcia-Fernandez18}, the track-oriented PMB
filter \cite{Williams15a} and the PMBM filter with intra-track Bernoulli
merging \cite{Fontana2020} against their clustered versions.

The filter implementations use a threshold for pruning the Poisson
components $\Gamma_{p}=10^{-5}$, a threshold for pruning global hypotheses
$\Gamma_{mbm}=10^{-4}$, and a threshold for pruning Bernoulli components
$\Gamma_{b}=10^{-5}$. The maximum number of global hypotheses is
$N_{h}=200$ for the standard PMBM and PMB filters, while the limit
on the number of global hypotheses is $N_{h}^{c}=20|C^{c}|$ for each
cluster $c$ in the clustered versions of the filters. The ellipsoidal
gating is performed with a $k$-d tree of threshold $\Gamma_{g}=4.5\sigma_{k}^{i,a^{i}}$,
while the estimation is performed selecting the global hypothesis
with the highest weight and reporting Bernoulli components whose existence
probability is above $0.4$ \cite[Sec. VI.A]{Garcia-Fernandez18}.
The intra-track Bernoulli merging procedure has threshold $\Gamma_{m}=0.25$
to determine similar Bernoulli components, and the inter-track Bernoulli
has its threshold set at $\Gamma_{s}=50$. These parameters have been
determined empirically for good performance, and they represent a
reasonable trade-off between computational burden and accuracy. Note
that the clustering and merging methods only add two parameters to
those required by the standard PMBM filter implementation.

Furthermore, we provide a comparison with the efficient implementation
of the $\delta$-GLMB\cite{Vo2014} with joint prediction and update, with
and without adaptive birth. The number of global hypotheses and birth
model in $\delta$-GLMB are matched with the standard PMBM filter
\cite{GarciaFernandez2020a}. All filters have been implemented using
Murty's algorithm \cite{Murty1968}.

In the simulations, target motion follows a nearly constant velocity
model \cite{BarShalom2001}. The target state is described in a two-dimensional
Cartesian coordinate system by $s_{k}=[p_{x,k},v_{x,k},p_{y,k},v_{y,k}]^{T}$,
where the first two components represent position and velocity of
the target on the $x$-axis, and the last two those on the $y$-axis.
The parameters of the linear and Gaussian motion and measurement models
are
\begin{align*}
F & =I_{2}\otimes\begin{pmatrix}1 & T\\
0 & 1
\end{pmatrix}, & Q & =qI_{2}\otimes\begin{pmatrix}T^{3}/3 & T^{2}/2\\
T^{2}/2 & T
\end{pmatrix}
\end{align*}
\[
H=I_{2}\otimes\begin{pmatrix}1 & 0\end{pmatrix},\quad R=I_{2}
\]
where $\otimes$ is the Kronecker product, $T=1$ is the sampling
period, and $q=0.01$ or $q=0.2$ in Scenario 1 and 2, respectively.
The clutter model is Poisson, uniformly distributed in the area of
interest, with a mean number of clutter measurements per scan $\lambda_{c}$
dependent on the area of interest in each scenario. We set the probability
of survival of the targets $p_{S}=0.99$, and the probability of detection
$p_{D}=0.9$ for all the simulations. To evaluate the performance
of the algorithm, we consider the root mean square (RMS) of the GOSPA
error ($\alpha=2$, $c=10$, $p=2$) \cite{Rahmathullah17}, which
allows us to decompose the total error into localization error, missed
target error and false target error.

\begin{figure}
\centering{}\includegraphics[scale=0.35]{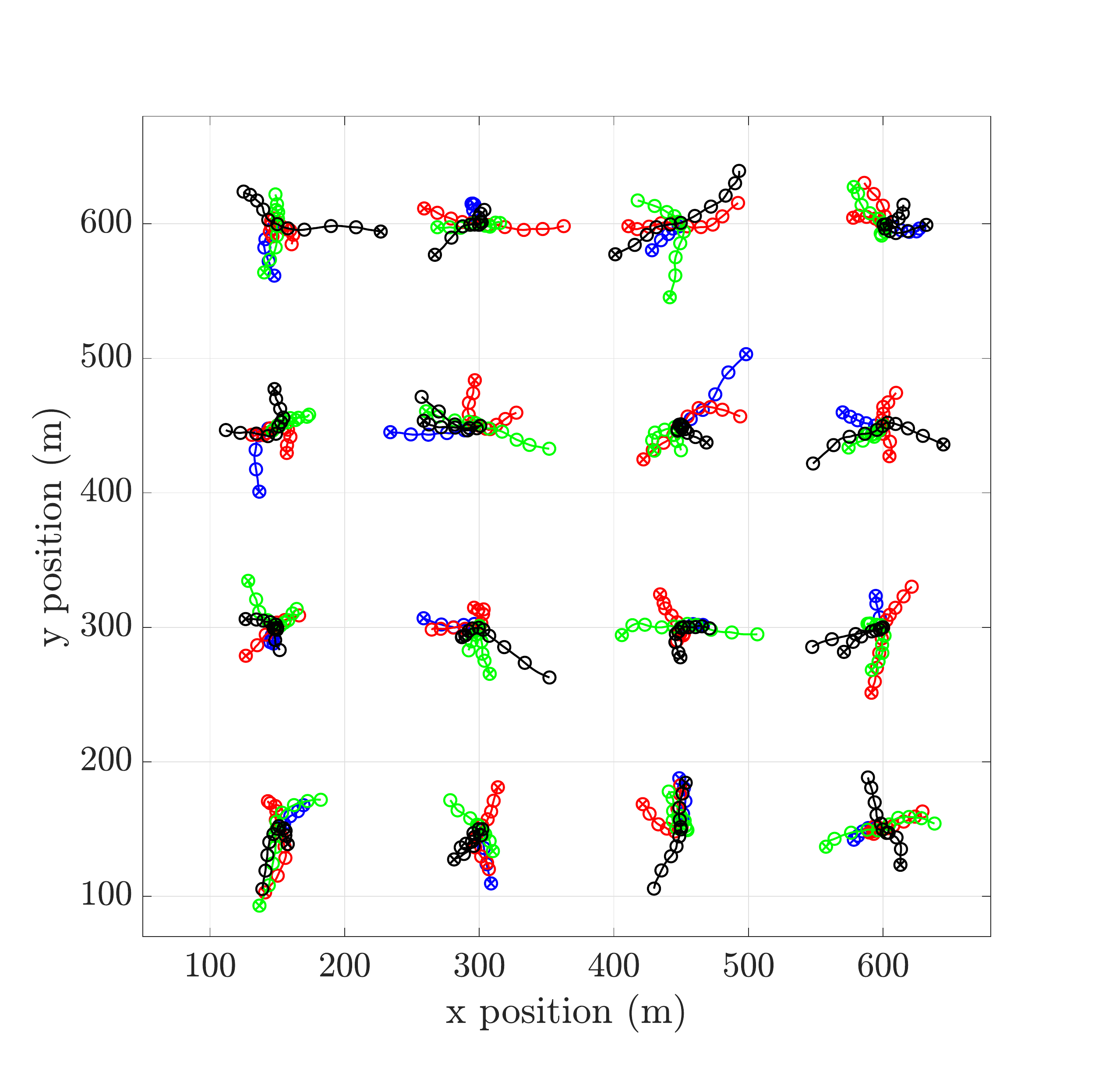}\caption{Example of a simulation $N_{sim}=2$ in Scenario 1. Each group is
composed by four targets, and each target is depicted with a different
colour within the group. All the targets are born at time step $k=1$
and survive for 101 time steps, except the blue targets that die at
time step $k=50$ in all the groups. The target positions at $k=1$
are indicated by a cross, and the circles show the target positions
every ten time steps.\label{fig:Extended_scenario}}
\end{figure}

\begin{figure}
\centering{}\includegraphics[scale=0.35]{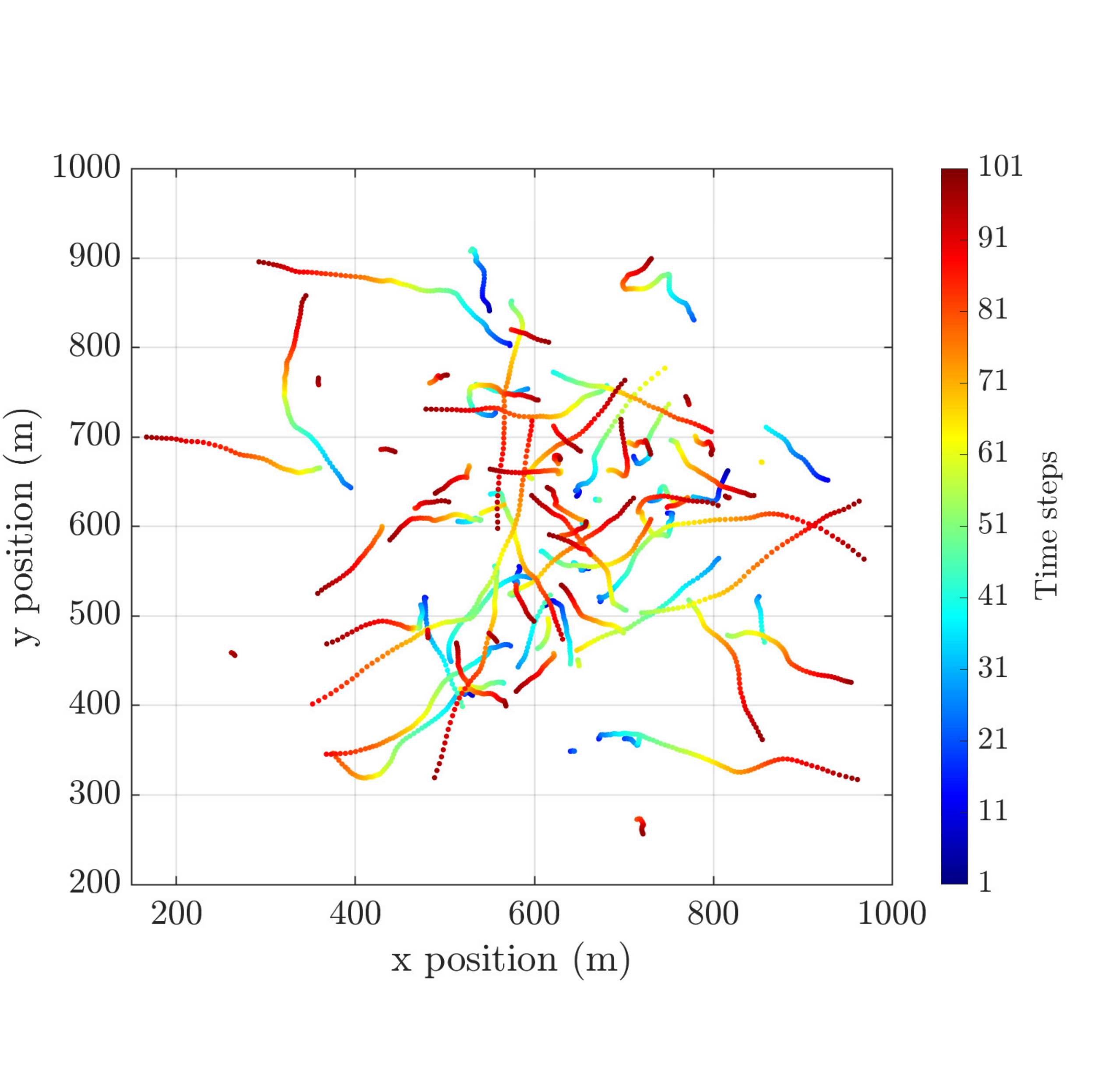}\caption{Example of a simulation $N_{sim}=2$ in Scenario 2. The colours represent
the evolution over time of the target positions in the field of view.\label{fig:Extended_scenario2}}
\end{figure}

We consider two scenarios based on different parameters and structure,
as shown in Fig. \ref{fig:Extended_scenario} and \ref{fig:Extended_scenario2}.
For each scenario, we perform four simulations denoted by the index
$N_{sim}$ and defined by the number of groups of targets $N_{g}$,
the mean number of targets born during the simulation $N_{b}$, the
mean number of targets alive at each time step $N_{a}$, the side
length of the area of interest $d_{A}$ and the mean number of clutter
measurements per scan $\lambda_{c}$. All units in this section are
expressed in the international system and omitted for notational clarity.
Tab. \ref{tab:Sim_parameters} reports the parameters of the simulations
and the mean number of global hypotheses before and after pruning,
$N_{GH}^{b}$ and $N_{GH}^{a}$, respectively, in the standard PMBM
for each simulation and scenario.

The simulations have been performed on a laptop equipped with Intel
(R) Core(TM) i7-8850H @ 2.60 GHz and 16 GB of memory. All the codes
are written in MATLAB, except for Murty\textquoteright s algorithm
and R-Tree, which are written in C++\footnote{We used the Murty's algorithm implementation in the tracker component
library \cite{Crouse2017}, and a modification of the R-Tree algorithm
by Antonin Guttman available on https://github.com/nushoin/RTree. }, and the priority queue, which is based on a Python implementation.
The results are based on the average on 50 Monte Carlo (MC) runs,
except for those related to the simulations $N_{sim}=4$, which are based
on 30 MC runs due to the long execution timesfor the standard PMBM
filter.

Note that all filters use sub-optimal estimators and different approximations,
like pruning, merging and clustering. While the PMBM filter without
approximations and an optimal estimator provides optimal estimates
of the set of targets, a PMB filter implementation can perform better
than a PMBM filter implementation with a sub-optimal estimator and
approximations.

\begin{table}
\centering{}\caption{Simulations parameters for Scenario 1 and 2. The number of groups
of targets $N_{g}$ is not defined in scenario 2, as the targets are
born in the same area of interest. \label{tab:Sim_parameters}}
\begin{tabular}{ccccc|cccc}
\hline 
\noalign{\vskip\doublerulesep}
 &
\multicolumn{4}{c}{Scenario 1} &
\multicolumn{4}{c}{Scenario 2}\tabularnewline
\noalign{\vskip\doublerulesep}
$N_{sim}$ &
1 &
2 &
3 &
4 &
1 &
2 &
3 &
4\tabularnewline
\hline 
\hline 
\noalign{\vskip\doublerulesep}
$N_{g}$ &
4 &
16 &
64 &
256 &
N/D &
N/D &
N/D &
N/D\tabularnewline
\noalign{\vskip\doublerulesep}
$N_{b}$ &
16 &
64 &
256 &
1024 &
16 &
64 &
256 &
1024\tabularnewline
\noalign{\vskip\doublerulesep}
$N_{a}$ &
14 &
56 &
224 &
895 &
6 &
24 &
96 &
374\tabularnewline
\noalign{\vskip\doublerulesep}
$d_{A}$ &
400 &
750 &
1350 &
2550 &
600 &
1200 &
1800 &
2400\tabularnewline
\noalign{\vskip\doublerulesep}
$\lambda_{c}$ &
2.25 &
6.25 &
20.25 &
72.25 &
24 &
96 &
216 &
384\tabularnewline
\noalign{\vskip\doublerulesep}
$N_{GH}^{b}$ &
267 &
284 &
300 &
296 &
143 &
242 &
258 &
259\tabularnewline
\noalign{\vskip\doublerulesep}
$N_{GH}^{a}$ &
126 &
163 &
191 &
\multicolumn{1}{c}{195} &
33 &
61 &
88 &
91\tabularnewline[\doublerulesep]
\hline 
\end{tabular}
\end{table}

\subsection{Gating\label{subsec:Gating-1}}

\begin{figure}
\centering{}\includegraphics[scale=0.32]{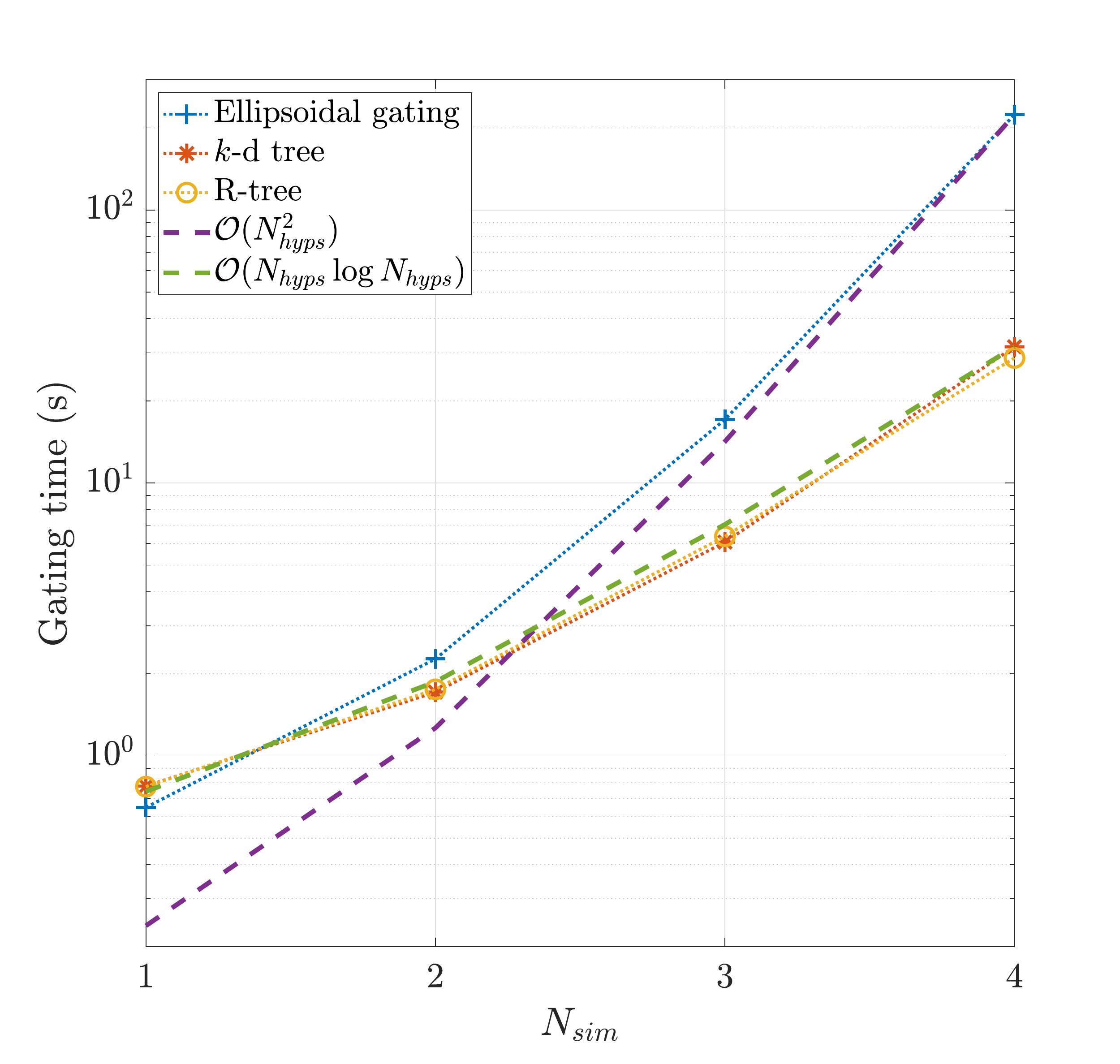}\caption{Comparison between the computational time of the gating procedure
using different data structures (5MC runs). The dashed lines represents
the asymptotic computational complexity based on $N_{hyps}$. \label{fig:Gating_times_comp}}
\end{figure}

In Fig. \ref{fig:Gating_times_comp}, we compare the mean gating times
of several gating procedures. The gating time is defined as the time
to update the single target hypotheses in the prediction density $\widetilde{f}_{k|k-1}\left(\cdot\right)$,
and it includes the time to build and query the space partitioning
data structure. It also considers the time to generate the misdetection
hypotheses and to compute the expected target measurements and the
innovation covariances. The gating thresholds for the ellipsoidal,
$k$-d tree and R-Tree gating are $\gamma_{G}=20$, $\gamma_{G}=4.5$
and $\gamma_{G}=8$ respectively, and they provide equivalent results.

Fig. \ref{fig:Gating_times_comp} highlights the asymptotic computational
complexity based on the mean of the total number of single target
hypotheses $N_{hyps}$ generated throughout all the time instants
of each simulation. The use of $k$-d trees or R-trees has a computational
burden associated with the initialisation step, which overcomes the
benefits of using data structures for $N_{sim}=1$. This computational
effort is rewarded with faster queries, resulting in a relevant speed-up
of the gating procedure as the number of targets increases in the
simulations. The comparison between $k$-d tree and R-Tree yields
a limited difference in terms of gating time, and results independent
on the number of targets.

\subsection{Scenario 1\label{subsec:Scenario-1}}

Scenario 1 is an extension of the base scenario proposed in \cite{Garcia-Fernandez18},
which consists of four targets, all born at time step 1 and alive
throughout the simulation of 101 time steps, except one which dies
at time step 50 (the blue ones in Fig. \ref{fig:Extended_scenario}).
The base scenario is considered challenging, as all the targets get
close at time step 50, when the blue one dies. We extended the base
scenario by generating $N_{g}$ groups of four targets in the area
of interest, as shown in Fig. \ref{fig:Extended_scenario}. Each group
of four targets is generated at a distance $d_{offset}=150$ from
the centre of the adjacent groups, within a square area of side length
$d_{a}=300$. The total area of interest is $A=[0,d_{A}(N_{g})]\times[0,d_{A}(N_{g})]$,
where $d_{A}(N_{g})=d_{a}+d_{offset}\cdot(N_{g}-1)$.

We test the scenario with four configurations based on different numbers
of groups $N_{g}$ as indicated in Tab. \ref{tab:Sim_parameters}.
In each configuration, the targets are born according to a PPP of
intensity $\lambda\cdot u_{A}(z)$ at the first time step, where $u_{A}(z)$
is a uniform density in its area of interest and $\lambda=3N_{g}$;
the PPP intensity decrease to $0.005$ at the next time steps. The
intensity is Gaussian with mean $[d_{A}(N_{g})/2,0,d_{A}(N_{g})/2,0]^{T}$,
and covariance diag$([(1.1d_{A}(N_{g}))^{2},1,(1.1d_{A}(N_{g}))^{2},1])$.

\subsubsection{Clustering\label{subsec:Clustering-1}}

\begin{figure}
\centering{}\includegraphics[scale=0.32]{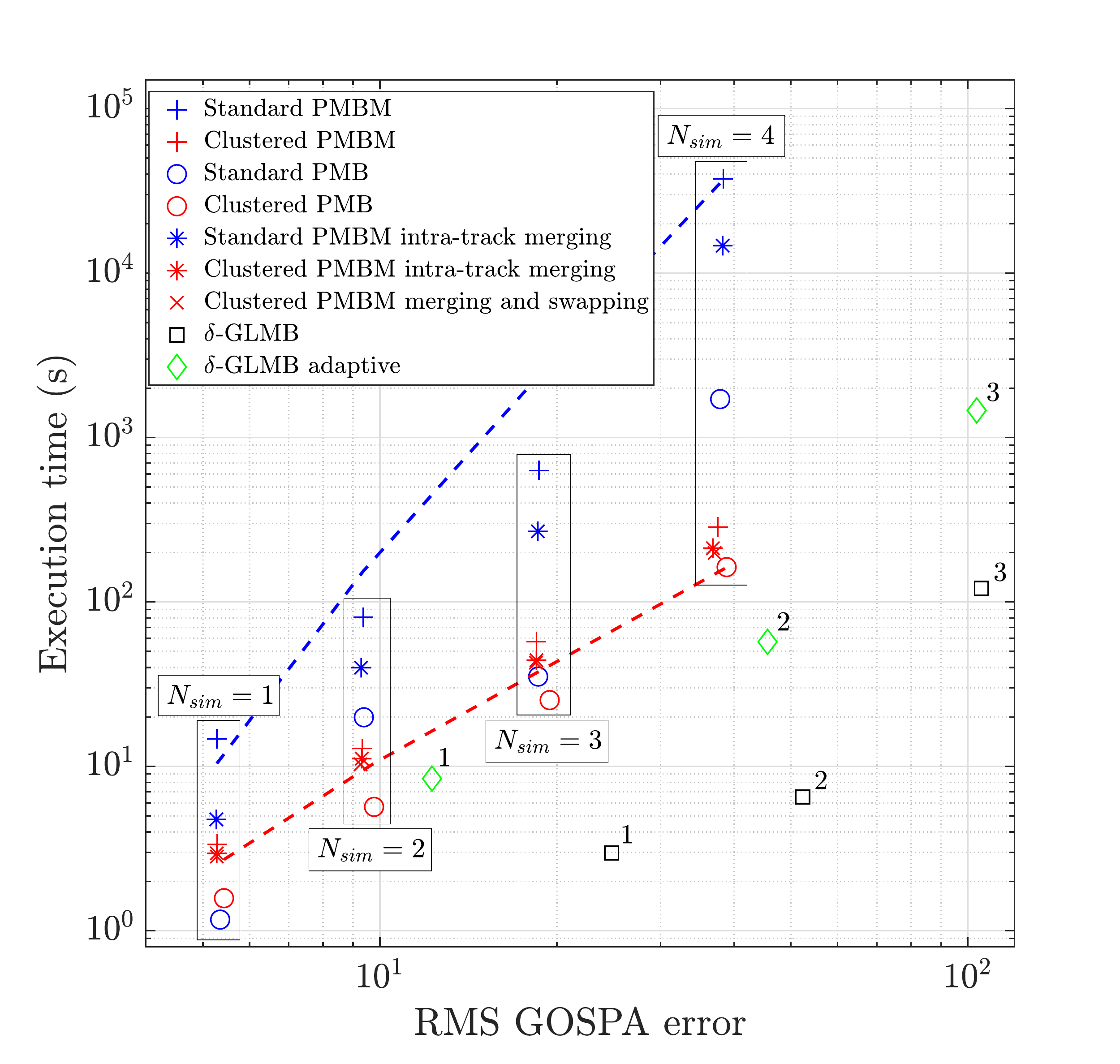}\caption{\label{fig:Sim_diff_number_targetsScenario1}Comparison between performance
and execution times between the standard PMBM and PMB filters (in
blue),their clustered versions (in red) and $\delta$-GLMB  in Scenario
1. Each version of the filter is represented with a different marker.
The simulations $N_{sim}=[1,2,3,4]$ have mean number of targets born
during the simulation $N_{b}=[16,64,256,1024]$, and they indicated
with a box or a superscript number ($\delta$-GLMB). The blue and
red dashed lines represent the asymptotic computational complexity
$\mathcal{O}(N_{a}^{2})$ and $\mathcal{O}(N_{a})$, respectively.}
\end{figure}

In Fig. \ref{fig:Sim_diff_number_targetsScenario1} the results of
the simulations based on Scenario 1 are indicated using different
markers, and the asymptotic computational complexity based on the
mean number of targets alive at each time step $N_{a}$ is expressed
by the two dashed lines. Note that the computational gains are highly
dependent on the scenario, and we cannot draw general conclusions. 

The outcome shows reduced computational time for both the clustered
PMBM and PMB compared to their standard implementations and the $\delta$-GLMB
filter with adaptive birth. Notably, the clustered PMBM and its variations
based on Bernoulli merging and swapping show the same performance
than the standard filter, where the two Bernoulli reduction techniques
provide even lower execution times, as indicated in more detail in
Tab. \ref{tab:Performance_inter_track_merge}. The $\delta$-GLMB
filter without adaptive birth results faster than PMBM, but it shows
a high GOSPA error, especially in the misdetected target error, due
the great number of targets born simultaneously at the beginning of
the simulations.

The clustered PMB filter is usually faster than the correspondent
standard implementation, although it results less accurate. The management
cost of the clusters overcomes the benefits of our approach in scenarios
with a low number of targets, e.g. $N_{sim}=1$. Similar conclusions
can be drawn for $p_{D}=0.7$, see App. \ref{sec:Evaluation-with-low}.

\begin{table*}
\centering{}\caption{Performance and computation time of the clustered PMBM with Bernoulli
merging and swapping techniques based on different simulations in
Scenario 1. The simulations $N_{sim}=[1,2,3,4]$ have mean number
of targets born during the simulation $N_{b}=[16,64,256,1024]$. \label{tab:Performance_inter_track_merge}}
\begin{tabular}{ccccc|cccc}
\hline 
\noalign{\vskip\doublerulesep}
 &
\multicolumn{4}{c}{RMS GOSPA error} &
\multicolumn{4}{c}{Time (s)}\tabularnewline
\noalign{\vskip\doublerulesep}
$N_{sim}$ &
$1$ &
$2$ &
$3$ &
$4$ &
$1$ &
$2$ &
$3$ &
$4$\tabularnewline
\hline 
\hline 
\noalign{\vskip\doublerulesep}
Standard PMBM &
5.28 &
9.37 &
18.65 &
38.35 &
14.71 &
80.78 &
630.05 &
37523.1\tabularnewline
\noalign{\vskip\doublerulesep}
Clustered PMBM &
5.29 &
9.33 &
18.46 &
37.09 &
3.36 &
12.86 &
57.33 &
285.61\tabularnewline
\noalign{\vskip\doublerulesep}
Clustered PMBM intra-track merging &
5.28 &
9.31 &
18.46 &
36.84 &
2.97 &
11.16 &
44.25 &
212.50\tabularnewline
\noalign{\vskip\doublerulesep}
Clustered PMBM merging and swapping &
5.27 &
9.27 &
18.42 &
37.08 &
2.83 &
10.28 &
42.69 &
198.73\tabularnewline
\hline 
\end{tabular}
\end{table*}

\subsubsection{Inter-track Bernoulli merging\label{subsec:Inter-track-Bernoulli-merging}}

\begin{figure}
\begin{centering}
\par\end{centering}
\centering{}\includegraphics[scale=0.17]{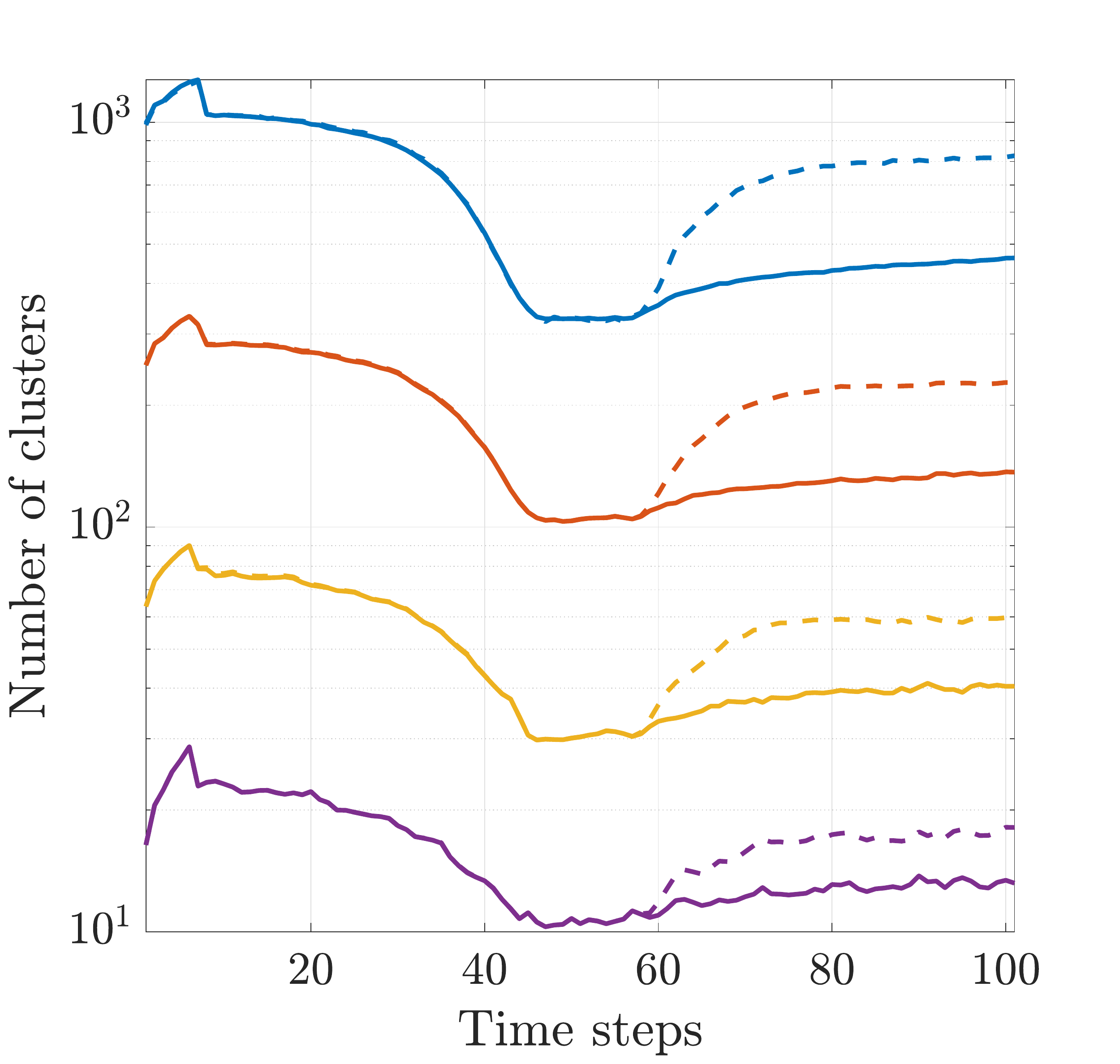}\includegraphics[scale=0.17]{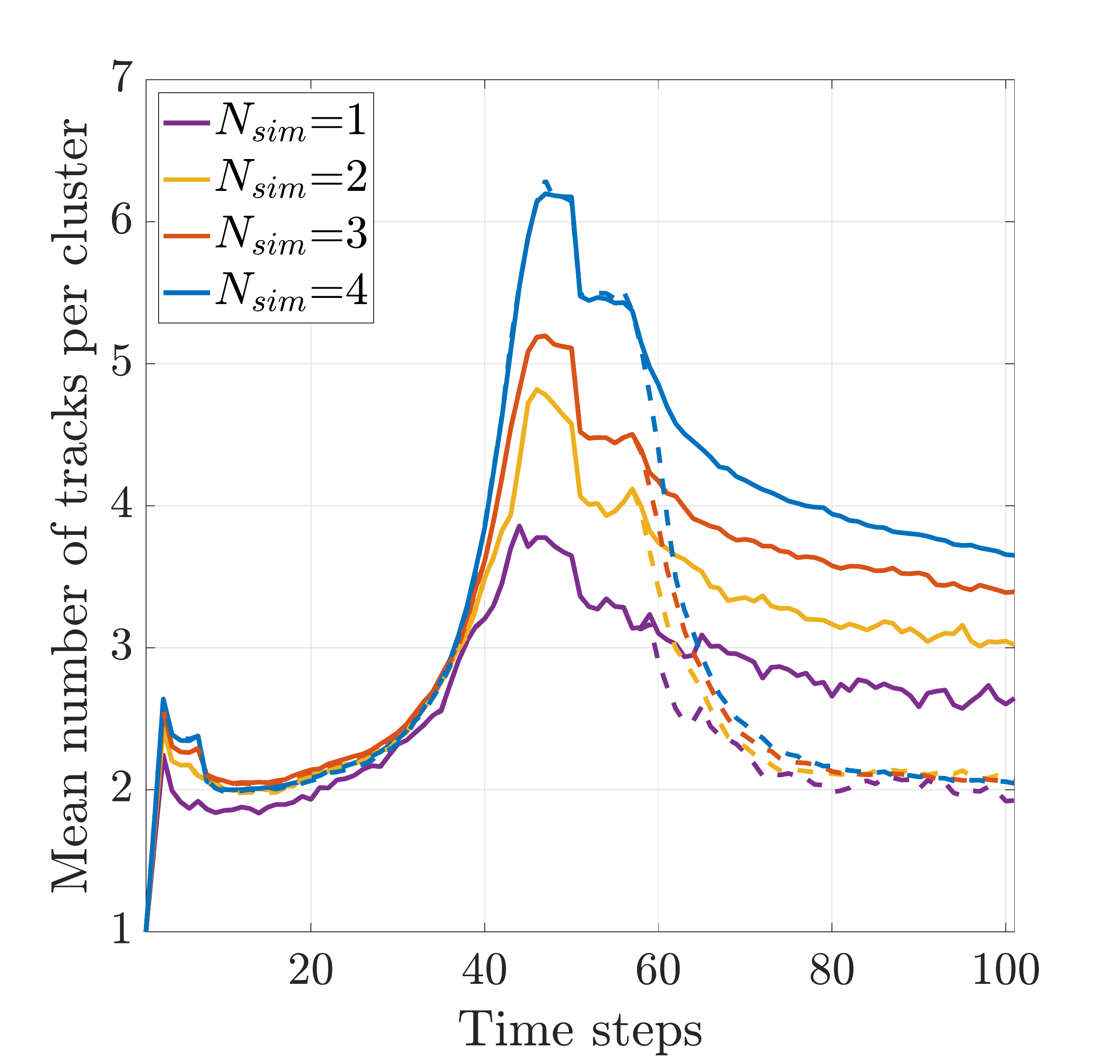}\caption{\label{fig:Statistics_inter_track_merge}Comparison of the mean number
of clusters and mean number of tracks per cluster for the clustered
PMBM with intra-track and inter-track Bernoulli merging. The simulations
are based on different simulations in Scenario 1. Solid lines correspond
to the clustered PMBM performing only intra-track merging, while dashed
lines correspond to the clustered PMBM performing both the intra and
inter track Bernoulli merging and swapping procedures. }
\end{figure}

Tab. \ref{tab:Performance_inter_track_merge} compares the results
of the simulations based on Scenario 1 of the clustered PMBM filter
approximated with the intra and inter track Bernoulli merging and
swapping procedures. As already noticed, these techniques presents
a greater reduction of the computational time in the most challenging
scenarios, e.g. $N_{sim}=\{3,4\}$. This observation is supported
by the statistics reported in Fig. \ref{fig:Statistics_inter_track_merge},
where it is possible to notice a significant increase in the number
of clusters using the inter-track Bernoulli swapping procedure. Moreover,
the mean number of tracks per cluster falls to two regardless of the
number of targets in the simulation, which suggests the formation
of efficient clusters comprising an updated track and a new one related
by the same measurement.

\subsection{Scenario 2\label{subsec:Scenario-2}}

Scenario 2 considers targets that appear and disappear at different
time instants in an area $A=[0,d_{A}(N_{sim})]\times[0,d_{A}(N_{sim})]$,
$d_{A}=600N_{sim}$. The target state at the appearing time is Gaussian
with mean $[d_{A}(N_{sim})/2,0,d_{A}(N_{sim})/2,0]^{T}$ and covariance
diag$([(60N_{sim})^{2},1,(60N_{sim})^{2},1])$. The probability of
survival is $p_{D}=0.99$ and the expected number of targets born
at each time step is $N_{b}/100$, where $N_{b}$ is indicated in
Tab. \ref{tab:Sim_parameters}. The PPP intensity is Gaussian with
mean $[d_{A}(N_{sim})/2,0,d_{A}(N_{sim})/2,0]^{T}$, and covariance
diag$([(1.1d_{A}(N_{sim}))^{2},1,(1.1d_{A}(N_{sim}))^{2},1])$.

\begin{figure}
\centering{}\includegraphics[scale=0.32]{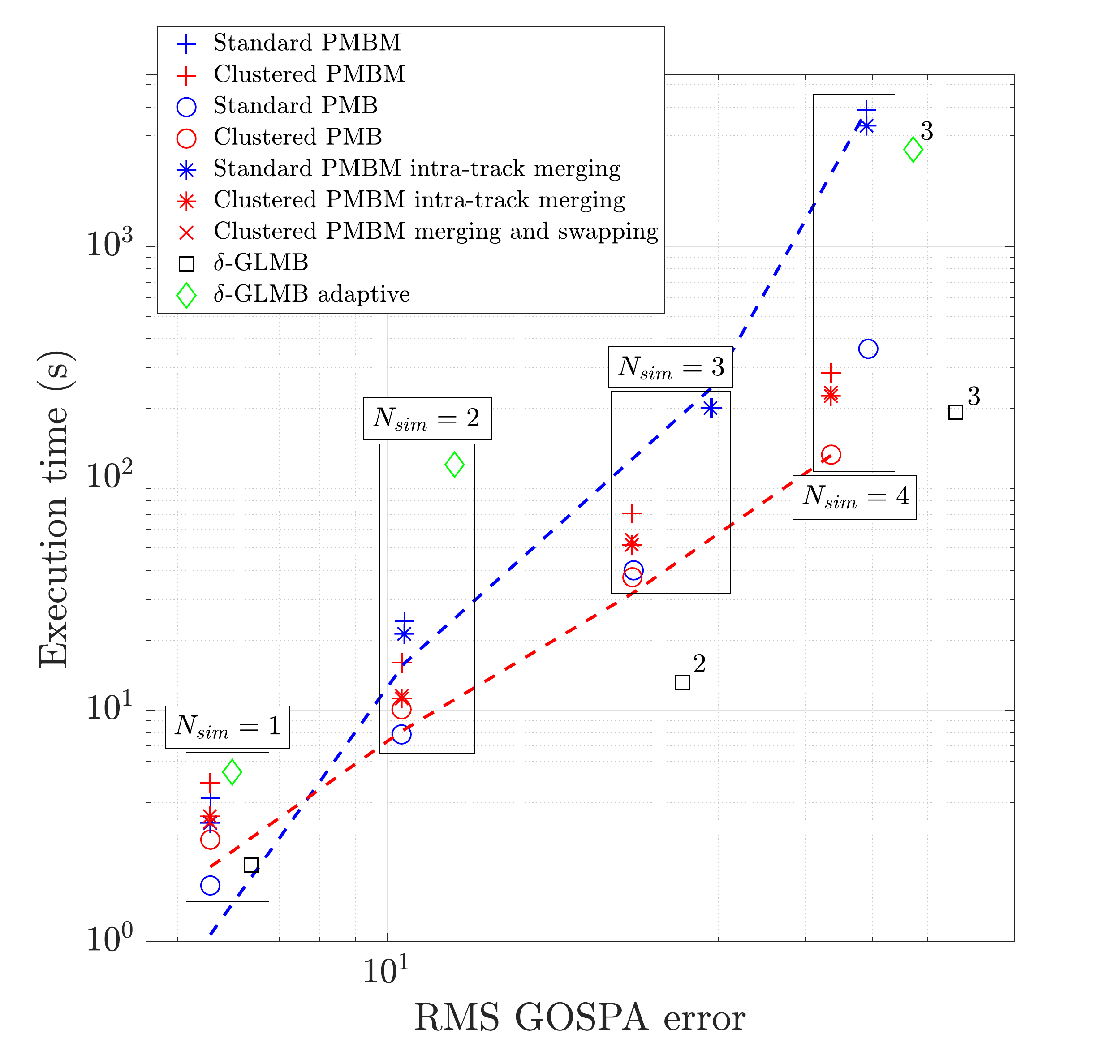}\caption{\label{fig:Sim_diff_number_targetsScenario2}Comparison between performance
and execution times between the standard PMBM and PMB filters (in
blue), their clustered versions (in red) and $\delta$-GLMB in Scenario
2. Each version of the filter is represented with a different marker.
The simulations $N_{sim}=[1,2,3,4]$ have mean number of targets born
during the simulation $N_{b}=[16,64,256,1024]$, and they indicated
with a box or a superscript number ($\delta$-GLMB). The blue and
red dashed lines represent the asymptotic computational complexity
$\mathcal{O}(N_{a}^{2})$ and $\mathcal{O}(N_{a})$, respectively.}
\end{figure}

In Fig. \ref{fig:Sim_diff_number_targetsScenario2} we show the results
of the simulations based on Scenario 2. The RMS GOSPA error of the
PMBM filter results higher than the respective clustered version in
simulations $N_{sim}\in\{3,4\}$. The reason is related to the high
clutter rate in these simulations, which generates a significant number
of global hypotheses. Most of these hypotheses are pruned in the PMBM
filter due to the cap on the maximum number of global hypotheses.
The distributed representation of the global hypotheses implemented
by the clustered PMBM allows us to express a higher number of global
hypotheses in a more efficient way, resulting in better performance
in reduced computational time. Note that, as in Scenario 1, if the
number of targets is low, the management cost of the clusters overcomes
the benefits of our approach for some filters.

\section{Conclusions\label{sec:Conclusions}}

In this paper, we have proposed several algorithms to reduce the complexity
of the PMBM filter, enabling its use in complex scenarios with high
number of targets. We have introduced the clustered PMBM density and
a clustering algorithm based on the measurements associated with the
potential targets. We have also proposed two techniques to decrease
the number of similar Bernoulli components that arise while filtering,
namely intra-track Bernoulli merging and inter-track Bernoulli swapping.

We have evaluated the filters in two simulated scenarios showing the
advantages of PMBM clustering for a high number of targets. Considering
clustering, tracking well-separated targets splits the problem into
different targets reaching linear complexity with a clustered PMBM
filter. If all targets are in close proximity, it is not possible
to split targets into clusters, and the PMBM and clustered PMBM should
have similar complexity.

Future work is the development of a hybrid MPI (Message Passing Interface)
and OpenMP (Open Multi-Processing) \cite{Rabenseifner2006} implementation,
such that the algorithm can be scaled beyond the capabilities of the
shared memory systems for which OpenMP alone is applicable. Future
lines of research can also include the use of the proposed merging
strategies to perform multi-sensor average fusion in the PMBM/PMB
framework \cite{Li2023}. Furthermore, other clustering techniques
can be explored to perform clustered PMBM filtering.\bibliographystyle{IEEEtran}
\bibliography{11C__Users_marfon_OneDrive_-_The_University_of_Liverpool_Marco_Fontana}

\cleardoublepage{}

{\LARGE{}Supplementary material: Data-driven clustering and Bernoulli
merging for the Poisson multi-Bernoulli mixture filter}{\LARGE\par}

\appendices{}

\section{Proof of Lemma \ref{lem:KLD_min}\label{sec:Proof-of-Lemma2}}

In this appendix, we prove Lemma \ref{lem:KLD_min}. Applying the
KLD in (\ref{eq:KLD_f_q}) can be written as \cite[ Eq. (3.53)]{Mahler14} 

\begin{align*}
\mathrm{D} & \left(\widetilde{f}_{k'|k}\right.\left\Vert \widetilde{q}_{k'|k}\right)=\int\widetilde{f}_{k'|k}\left(\widetilde{Y}_{k'}\uplus\widetilde{X}_{k'}^{1}\uplus...\uplus\widetilde{X}_{k'}^{n_{k'|k}}\right)\\
 & \times\log\frac{\widetilde{f}_{k'|k}\left(\widetilde{Y}_{k'}\uplus\widetilde{X}_{k'}^{1}\uplus...\uplus\widetilde{X}_{k'}^{n_{k'|k}}\right)}{\widetilde{q}_{k'|k}\left(\widetilde{Y}_{k'}\uplus\widetilde{X}_{k'}^{1}\uplus...\uplus\widetilde{X}_{k'}^{n_{k'|k}}\right)}\delta\widetilde{Y}_{k'}\delta\widetilde{X}_{k'}^{1}....\delta\widetilde{X}_{k'}^{n_{k'|k}}\\
 & =constant-\int\widetilde{f}_{k'|k}^{p}\left(\widetilde{Y}_{k'}\right)\log\widetilde{q}_{k'|k}^{0}\left(\widetilde{Y}_{k'}\right)\delta\widetilde{Y}_{k'}\\
 & -\int\sum_{a\in\mathcal{A}_{k'|k}}w_{k'|k}^{a}\prod_{i=1}^{n_{k'|k}}\left[\widetilde{f}_{k'|k}^{i,a^{i}}\left(\widetilde{X}_{k'}^{i}\right)\right]\\
 & \times\log\left(\prod_{c=1}^{n_{k'|k}^{c}}\widetilde{q}_{k'|k}^{c}\left(\cup_{i\in C_{k}^{c}}\widetilde{X}_{k'}^{i}\right)\right)\delta\widetilde{X}_{k'}^{1}....\delta\widetilde{X}_{k'}^{n_{k'|k}}\\
 & =constant-\int\widetilde{f}_{k'|k}^{p}\left(\widetilde{Y}_{k'}\right)\log\widetilde{q}_{k'|k}^{0}\left(\widetilde{Y}_{k'}\right)\delta\widetilde{Y}_{k'}\\
 & -\sum_{c=1}^{n_{k'|k}^{c}}\sum_{a\in\mathcal{A}_{k'|k}}w_{k'|k}^{a}\int\prod_{i\in C_{k}^{c}}\left[\widetilde{f}_{k'|k}^{i,a^{i}}\left(\widetilde{X}_{k'}^{i}\right)\right]\\
 & \times\log\left(\widetilde{q}_{k'|k}^{c}\left(\cup_{i\in C_{k}^{c}}\widetilde{X}_{k'}^{i}\right)\right)\left[\prod_{i\in C^{c}}\delta\widetilde{X}_{k'}^{i}\right]
\end{align*}
where constant denotes terms that do not depend on $q$. By standard
KLD minimisation, we prove that the density on the augmented set of
undetected targets in the clustered density is equal to the PPP of
the PMBM density (\ref{eq:PPP_augmented}) with auxiliary variables.
Furthermore, the cluster density $\widetilde{q}_{k'|k}^{c}$ on the
augmented set of targets $C_{k}^{c}$ is proportional to the MBM expressing
the global hypotheses based on the Bernoulli components of the tracks
in the cluster.

\section{Proof of Lemma \ref{lem:IntOutAuxVar}\label{sec:Proof-of-Lemma3}}

In this appendix, we prove Lemma \ref{lem:IntOutAuxVar}, which provides
the relation between the clustered density of the clustered density
$q_{k'|k}(\cdot)$ and the clustered density $\widetilde{q}_{k'|k}(\cdot)$
with auxiliary variables.

\subsection{Preliminary result\label{subsec:Preliminary-result}}

Given a set with auxiliary variables, we can map it to a set without
auxiliary variables with the mapping
\begin{align*}
h\left(\left\{ \left(u_{1},x_{1}\right),...,\left(u_{n},x_{n}\right)\right\} \right) & =\left\{ x_{1},...,x_{n}\right\} .
\end{align*}
This is equivalent to a transition density
\begin{align}
f & \left(X_{k'}|\widetilde{X}_{k'}\right)=\delta_{h\left(\widetilde{X}_{k'}\right)}\left(X_{k'}\right)\nonumber \\
 & =\sum_{\uplus_{l=1}^{n_{k'|k}^{c}}X^{l}\uplus Y^{0}=X_{k'}}\delta_{h\left(\widetilde{Y}_{k'}\right)}\left(Y^{0}\right)\prod_{c=1}^{n_{k'|k}^{c}}\delta_{h\left(\cup_{i\in C_{k}^{c}}\widetilde{X}_{k'}^{i}\right)}\left(X^{c}\right)\label{eq:trans_density}
\end{align}
where $\delta_{h\left(\widetilde{X}_{k'}\right)}\left(\cdot\right)$
is the multi-target Dirac delta \cite{Mahler14}, and we have applied
that the multi-target Dirac delta can be seen as the union of independent
sets. 

We proceed to prove that we can recover (\ref{eq:integating_out_variables})
in Lemma \ref{lem:IntOutAuxVar}, by applying the transition density
$f\left(X_{k'}|\widetilde{X}_{k'}\right)$ in (\ref{eq:trans_density})
to a density $\widetilde{q}_{k'|k}\left(\cdot\right)$ and calculating
the set integral. As there is no change in cardinality the set integral
for fixed cardinality $n$ becomes
\begin{align*}
q(\left\{ x_{1},...,x_{n}\right\} )\\
=\frac{1}{n!}\sum_{u_{1:n}\in\mathbb{\mathbb{U}}_{k}^{n}} & \int f\left(\left\{ x_{1},...,x_{n}\right\} |\left\{ \left(u_{1},x'_{1}\right),...,\left(u_{n},x'_{n}\right)\right\} \right)\\
 & \times\widetilde{q}_{k'|k}\left(\left\{ \left(u_{1},x'_{1}\right),...,\left(u_{n},x'_{n}\right)\right\} \right)dx'_{1:n}\\
= & \frac{1}{n!}\sum_{u_{1:n}\in\mathbb{\mathbb{U}}_{k}^{n}}\int\delta_{\left\{ x'_{1},...,x'_{n}\right\} }\left(\left\{ x_{1},...,x_{n}\right\} \right)\\
 & \times\widetilde{q}_{k'|k}\left(\left\{ \left(u_{1},x'_{1}\right),...,\left(u_{n},x'_{n}\right)\right\} \right)dx'_{1:n}\\
= & \sum_{u_{1:n}\in\mathbb{\mathbb{U}}_{k}^{n}}\widetilde{q}_{k'|k}\left(\left\{ \left(u_{1},x_{1}\right),...,\left(u_{n},x_{n}\right)\right\} \right)
\end{align*}
which is equivalent to (\ref{eq:integating_out_variables}) in Lemma
\ref{lem:IntOutAuxVar}. Therefore, in the next subsection we prove
(\ref{eq:integating_out_variables}) in Lemma \ref{lem:IntOutAuxVar}
by applying the transition density $f\left(\cdot|\cdot\right)$ to
$\widetilde{q}_{k'|k}\left(\cdot\right)$. 

\subsection{Proof\label{subsec:Proof}}

We can rewrite (\ref{eq:q_no_details_MBM}) by explicitly considering
the convolution sum over the independent sets (see Definition \ref{def:aux_density})
\begin{align*}
\widetilde{q}_{k'|k} & \left(\widetilde{X}_{k'}\right)\\
= & \sum_{\uplus_{l=1}^{n_{k'|k}}\widetilde{X}^{l}\uplus\widetilde{Y}=\widetilde{X}_{k'}}\widetilde{q}_{k'|k}^{0}\left(\widetilde{Y}\right)\prod_{c=1}^{n_{k'|k}^{c}}\widetilde{q}_{k'|k}^{c}\left(\widetilde{X}^{c}\right).
\end{align*}
As we shown in the previous subsection, the clustered PMBM density
in $\mathcal{F}\left(\mathbb{R}^{n_{x}}\right)$ an be recovered by
applying the set integral
\begin{align*}
q_{k'|k}\left(X_{k'}\right) & =\int f\left(X_{k'}|\widetilde{X}_{k'}\right)\widetilde{q}_{k'|k}\left(\widetilde{X}_{k'}\right)\delta\widetilde{X}_{k'}
\end{align*}
where $f\left(\cdot|\cdot\right)$ is given by (\ref{eq:trans_density}).
Applying Lemma 2 in \cite{Williams15} yields
\begin{align*}
q_{k'|k} & \left(X_{k'}\right)\\
 & =\int\int f\left(X_{k'}|\widetilde{Y}\uplus\widetilde{X}^{1}\uplus...\uplus\widetilde{X}^{c_{k'|k}}\right)\\
 & \quad\times\widetilde{q}_{k'|k}^{0}\left(\widetilde{Y}\right)\prod_{c=1}^{n_{k'|k}^{c}}\widetilde{q}_{k'|k}^{c}\left(\widetilde{X}^{c}\right)\delta\widetilde{Y}\delta\widetilde{X}^{1:c_{k'|k}}\\
 & =\sum_{Y^{0}\uplus X^{1}\uplus...\uplus X^{c_{k'|k}}=X_{k'}}\left[\int\delta_{h\left(\widetilde{Y}^{0}\right)}\left(Y^{0}\right)\widetilde{q}_{k'|k}^{0}\left(\widetilde{Y}^{0}\right)\delta\widetilde{Y}\right]\\
 & \quad\times\prod_{c=1}^{n_{k'|k}^{c}}\left[\int\delta_{h\left(\widetilde{X}^{c}\right)}\left(X^{c}\right)\widetilde{q}_{k'|k}^{c}\left(\widetilde{X}^{c}\right)\delta\widetilde{X}^{c}\right].
\end{align*}
Now, using the fact that
\begin{align*}
q_{0}^{c}\left(Y^{0}\right) & =\int\delta_{h\left(\widetilde{Y}^{0}\right)}\left(X_{k'}\right)\widetilde{q}_{k'|k}^{0}\left(\widetilde{Y}\right)\delta\widetilde{Y}\\
q_{k'|k}^{c}\left(X^{c}\right) & =\int\delta_{h\left(\widetilde{X}^{c}\right)}\left(X^{c}\right)\widetilde{q}_{k'|k}^{c}\left(\widetilde{X}^{c}\right)\delta\widetilde{X}^{c}
\end{align*}
we finish the proof of (\ref{eq:integating_out_variables}).
\begin{table}
\centering{}\caption{Mean number of global hypotheses in the standard PMBM before and after
pruning in Scenario 1 with probability of detection $p_{D}=0.7$.
\label{tab:Sim_parameters-pD07}}
\begin{tabular}{cccc}
\hline 
\noalign{\vskip\doublerulesep}
 &
\multicolumn{3}{c}{Scenario 1}\tabularnewline
\noalign{\vskip\doublerulesep}
$N_{sim}$ &
1 &
2 &
3\tabularnewline
\hline 
\noalign{\vskip\doublerulesep}
$N_{GH}^{b}$ &
286 &
300 &
305\tabularnewline
\noalign{\vskip\doublerulesep}
$N_{GH}^{a}$ &
42 &
175 &
183\tabularnewline[\doublerulesep]
\hline 
\end{tabular}
\end{table}

\section{Proof of Lemma \ref{lem:Recursive_clustered}\label{sec:Proof-of-Lemma4}}

In this appendix, we prove Lemma \ref{lem:Recursive_clustered}. Applying
the KLD in (\ref{eq:KLD_f_q}) and defining the density $\widetilde{q}_{k'|k-1}^{c'}\left(\cdot\right)$
as
\begin{align*}
\widetilde{q}_{k'|k}^{c'}\left(\widetilde{X}_{k'}^{i}\right) & =\begin{cases}
1-r_{k'|k}^{i,a^{i}} & \widetilde{X}_{k'}^{i}=\emptyset\\
r_{k'|k}^{i,a^{i}}p_{k'|k}^{i,a^{i}}(x)\delta_{i}[u] & \widetilde{X}_{k'}^{i}=\left\{ (u,x):u\in c'\right\} \\
0 & \mathrm{otherwise}
\end{cases}
\end{align*}
and $\widetilde{X}_{k'}=\widetilde{X}_{k'}^{1}\uplus...\uplus\widetilde{X}_{k'}^{n_{k'|k-1}}$
, the KLD can be written as \cite[ Eq. (3.53)]{Mahler14} 

\begin{align*}
\mathrm{D} & \left(\widetilde{f}_{k'|k-1}\right.\left\Vert \widetilde{q}_{k'|k-1}\right)=\int\widetilde{f}_{k'|k-1}\left(\widetilde{Y}_{k'}\uplus\widetilde{X}_{k'}\right)\\
 & \times\log\frac{\widetilde{f}_{k'|k-1}\left(\widetilde{Y}_{k'}\uplus\widetilde{X}_{k'}\right)}{\widetilde{q}_{k'|k-1}\left(\widetilde{Y}_{k'}\uplus\widetilde{X}_{k'}\right)}\delta\widetilde{Y}_{k'}\delta\widetilde{X}_{k'}\\
 & =constant-\int\widetilde{f}_{k'|k-1}^{0}\left(\widetilde{Y}_{k'}\right)\log\widetilde{q}_{k'|k-1}^{0}\left(\widetilde{Y}_{k'}\right)\delta\widetilde{Y}_{k'}\\
 & -\int\prod_{c=1}^{n_{k-1}^{c}}\sum_{a\in\mathcal{A}_{k'|k-1}}w_{k'|k-1}^{a}\prod_{i\in C_{k-1}^{c}}\left[\widetilde{f}_{k'|k-1}^{i,a^{i}}\left(\widetilde{X}_{k'}^{i}\right)\right]\\
 & \times\log\left(\prod_{c'=1}^{n_{k'|k}^{c'}}\widetilde{q}_{k'|k-1}^{c'}\left(\cup_{i\in C_{k}^{c'}}\widetilde{X}_{k'}^{i}\right)\right)\delta\widetilde{X}_{k'}^{1}...\delta\widetilde{X}_{k'}^{n_{k'|k}}\\
 & =constant-\int\widetilde{f}_{k'|k-1}^{0}\left(\widetilde{Y}_{k'}\right)\log\widetilde{q}_{k'|k-1}^{0}\left(\widetilde{Y}_{k'}\right)\delta\widetilde{Y}_{k'}\\
 & -\sum_{a\in\mathcal{A}_{k'|k}}w_{k'|k-1}^{a}\sum_{c=1}^{n_{k'|k}^{c}}\sum_{c'=1}^{n_{k'|k}^{c'}}\int\prod_{i\in C_{k}^{c}}\left[\widetilde{f}_{k'|k-1}^{i,a^{i}}\left(\widetilde{X}_{k'}^{i}\right)\right]\\
 & \times\log\left(\widetilde{q}_{k'|k-1}^{c'}\left(\cup_{i\in C_{k}^{c}}\widetilde{X}_{k'}^{i}\right)\right)\left[\prod_{i\in C^{c}}\delta\widetilde{X}_{k'}^{i}\right]
\end{align*}
as in App. \ref{sec:Proof-of-Lemma2}, the constant denotes terms
that do not depend on $q$. As the cluster density $\widetilde{q}_{k'|k-1}^{c'}(\cdot)$
on the target not belonging to the cluster $c'$ is zero, we can rewrite
the last equation considering just the target states in the intersection
between the the clusters $c$ and $c'$
\begin{align*}
\mathrm{D} & \left(\widetilde{f}_{k'|k-1}\right.\left\Vert \widetilde{q}_{k'|k-1}\right)=\\
 & =constant-\int\widetilde{f}_{k'|k-1}^{0}\left(\widetilde{Y}_{k'}\right)\log\widetilde{q}_{k'|k-1}^{0}\left(\widetilde{Y}_{k'}\right)\delta\widetilde{Y}_{k'}\\
 & -\sum_{a\in\mathcal{A}_{k'|k}}w_{k'|k-1}^{a}\sum_{c=1}^{n_{k'|k}^{c}}\sum_{c'=1}^{n_{k'|k}^{c'}}\int\prod_{i\in C^{c}\cap C^{c'}}\left[\widetilde{f}_{k'|k-1}^{i,a^{i}}\left(\widetilde{X}_{k'}^{i}\right)\right]\\
 & \times\log\left(\widetilde{q}_{k'|k-1}^{c'}\left(\cup_{i\in C^{c}\cap C^{c'}}\widetilde{X}_{k'}^{i}\right)\right)\left[\prod_{i\in C^{c}\cap C^{c'}}\delta\widetilde{X}_{k'}^{i}\right]
\end{align*}
which proves the proportionality between the cluster prediction density
$\widetilde{q}_{k'|k-1}^{c'}(\cdot)$ and the product of the MBMs
based on the Bernoulli components of the targets belonging to the
intersection $C^{c}\cap C^{c'}$, for $c\in\{1,\dots,n_{k-1}^{c}\}$.
\begin{figure}
\centering{}\includegraphics[scale=0.32]{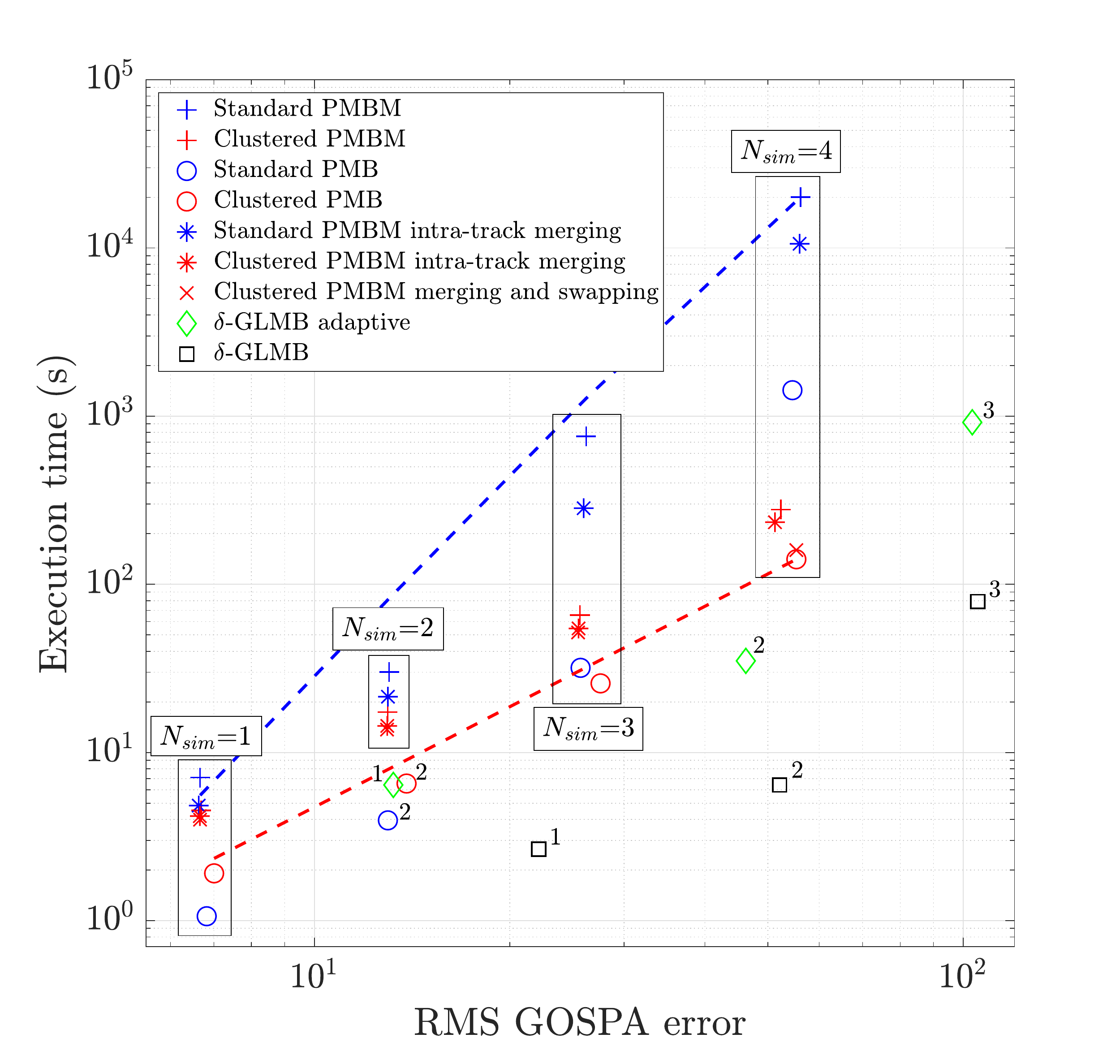}\caption{\label{fig:Sim_diff_number_targetsScenario1-pD07}Comparison between
performance and execution times between the standard PMBM and PMB
filters (in blue), their clustered versions (in red) and $\delta$-GLMB
in Scenario 1 with probability of detection $p_{D}=0.7$. Each version
of the filter is represented with a different marker. The simulations
$N_{sim}=[1,2,3,4]$ have mean number of targets born during the simulation
$N_{b}=[16,64,256,1024]$, and they indicated with a box or a superscript
number.}
\end{figure}

\section{An additional evaluation of the clustered PMBM filter\label{sec:Evaluation-with-low}}

In this appendix we present the evaluation of the clustered PMBM and
the proposed Bernoulli merging strategies in Scenario 1 with probability
of detection $p_{D}=0.7$. We provide a comparison with the standard
PMBM and PMB filters, and with the efficient implementation of the
$\delta$-GLMB with joint prediction and update, with and without
adaptive birth. The mean number of global hypotheses in simulations
$N_{sim}=\{1,2,3\}$ is reported in Tab. \ref{tab:Sim_parameters-pD07}.
The other simulation parameters are identical to those reported in
Sec. \ref{sec:Simulations}.

Fig. \ref{fig:Sim_diff_number_targetsScenario1-pD07} shows the performance
of the filters in terms of RMS GOSPA error and execution time. Compared
to the standard PMBM implementation, the proposed clustered methods
allow us to reduce the computational time providing similar accuracy.
As for higher $p_{D}=0.9$, the management cost of the clusters in
the PMB filter overcomes the benefits of our approach in scenarios
with a low number of targets ($N_{sim}=1$ and $N_{sim}=2$).

\end{document}